\documentclass[9pt]{IEEEtran}

\pdfoutput=1

\usepackage{cite}
\usepackage{amsmath,amssymb,amsfonts}
\usepackage{algorithmic}
\usepackage{etoolbox}
\usepackage{graphicx}
\usepackage{hyperref}
\usepackage{subcaption}
\usepackage{fancyhdr}

\begin{document}
	
\newbool{double_column}
\booltrue{double_column}

\title{Deep Learning Classification of 3.5 GHz Band Spectrograms with Applications to Spectrum Sensing} 

\author{
	\IEEEauthorblockN{
		W. Max Lees,
		Adam Wunderlich,
		Peter Jeavons,
		Paul D. Hale
		and Michael R. Souryal}
	\thanks{The authors are with the Communications Technology Laboratory, National Institute of Standards and Technology, Boulder, CO, USA and Gaithersburg, MD, USA (corresponding author e-mail: william.lees@nist.gov); 
		U.S. government work, not protected by U.S. copyright;
		Cleared for Open Publication  April 30, 2018 by the Department of Defense;
		Office of Prepublication and Security Review, Reference Number: 18-S-1288}}

\maketitle

\begin{abstract}
	In the United States, the Federal Communications Commission has adopted rules permitting commercial wireless networks to share spectrum with federal incumbents in the 3.5~GHz Citizens Broadband Radio Service band.  These rules require commercial systems to vacate the band when sensors detect radars operated by the U.S. military; a key example being the SPN-43 air traffic control radar.  Such sensors require highly-accurate detection algorithms to meet their operating requirements.  In this paper, using a library of over 14,000 3.5~GHz band spectrograms collected by a recent measurement campaign, we investigate the performance of thirteen methods for SPN-43 radar detection.  Namely, we compare classical methods from signal detection theory and machine learning to several deep learning architectures.  We demonstrate that machine learning algorithms appreciably outperform classical signal detection methods.  Specifically, we find that a three-layer convolutional neural network offers a superior tradeoff between accuracy and computational complexity.  Last, we apply this three-layer network to generate descriptive statistics for the full 3.5~GHz spectrogram library.  Our findings highlight potential weaknesses of classical methods and strengths of modern machine learning algorithms for radar detection in the 3.5~GHz band.
\end{abstract}

\section{Introduction}
	Spectrum sensing, defined as the task of ascertaining spectrum usage and the activity of primary users at a given location and time, is a key component of opportunistic spectrum usage by cognitive radio networks \cite{Yucek2009,Axell2012}.  Prior work on spectrum sensing algorithms has investigated methods from classical signal detection theory, such as energy detection and matched filtering \cite{Yucek2009,Axell2012}, classical machine learning techniques such as support vector machine (SVM) and K-nearest neighbor (KNN) classifiers \cite{Bkassiny2013,Thilina2013,Zhang2014,Lu2016,Jiang2017}, and distributed (or cooperative) sensing approaches involving multiple sensors \cite{Quan2008,Quan2009,Yucek2009,Axell2012,Thilina2013,Farrag2014,Lu2016}.  Recently, deep learning algorithms, such as  convolutional neural networks (CNNs) and long short-term memory (LSTM) recurrent neural networks \cite{Goodfellow2016}, which are the current state-of-the-art for many classification tasks \cite{LeCun2015,Goodfellow2016}, have been applied to various areas in wireless communications, including spectrum sensing, e.g., \cite{OShea2017,OShea2018,Rajendran2018,Zhang2018}.  

In a key step towards more efficient use of radio spectrum in the United States, the U. S. Federal Communications Commission (FCC) has adopted rules for the Citizens Broadband Radio Service (CBRS) that permit commercial wireless usage of the 3550-3700 MHz band (the ``3.5 GHz band") \cite{CFR-CBRS}.  The CBRS architecture outlined in the FCC rules includes a spectrum access system (SAS) together with environmental sensing capability (ESC) detectors to facilitate spectrum sharing in the 3550-3650 MHz band.  The purpose of the SAS is to coordinate commercial-user CBRS access so that federal incumbents are given priority access.  

The primary federal incumbents in the 3.5 GHz band are shipborne and ground-based radars operated by the U.S. military \cite{NTIAfasttrack2010}.  The CBRS framework requires that ESC sensors detect these radars, including the SPN-43 air traffic control radar \cite{SPN43manual}, also identified as Shipborne Radar 1 in \cite{NTIAfasttrack2010}.  ESC detection capabilities are determined by intended and unintended emissions, as well as background noise.  For example, out-of-band emissions (OOBE) from an adjacent-band U.S. Navy radar, identified as Shipborne Radar~3 in \cite{NTIAfasttrack2010}, are prevalent \cite{Cotton2014,Sanders2014,TN1954,TN1967}, and could complicate SPN-43 detection.  

As federal agencies collaborate with industry to refine standards and requirements for ESC detectors, sound methods for evaluation of ESC detector performance must be developed and potential limitations should be understood.  Furthermore, efforts to design ESC detectors could benefit from detection algorithm comparisons and the characterization of emissions in the 3.5 GHz band.  

In this paper, we address the above needs with a study of over 14,000 3.5 GHz band spectrograms recorded by a recent measurement campaign \cite{TN1954,TN1967} at two coastal locations: Point Loma, in San Diego, California and Fort Story, in Virginia Beach, Virginia.  Because the hardware required to record and process spectrograms is less complex, and therefore cheaper than that required for in-phase and quadrature (I/Q) data, the use of spectrograms by ESC sensors is an attractive option.  For this reason, our investigation focuses on SPN-43 detection from low-resolution spectrograms.

For the task of narrowband SPN-43 detection in a single 10~MHz channel observed by one receiver, we compare the effectiveness of thirteen detection algorithms, including eight deep learning methods, three classical machine learning approaches, and two strategies based on classical signal detection theory.  In addition, for the task of wideband SPN-43 detection across multiple channels observed concurrently with one receiver, we compare the top-performing methods from the single-channel evaluation.  A thorough performance evaluation utilizing a test set of unverified, human-labeled spectrograms reveals that deep learning methods outperform other approaches for SPN-43 detection.  Last, we apply the best-performing deep learning method to classify the complete set of spectrograms collected in San Diego and Virgina Beach with respect to SPN-43 presence, from which we estimate SPN-43 spectrum occupancy and characterize the power of non-SPN-43 emissions.

\section{3.5 GHz Spectrograms}
	\label{sec:spectrograms}
	As described in \cite{TN1954,TN1967}, 3.5~GHz band measurements were collected for a period of two months at each measurement site.  The primary aim of the measurements was to acquire high-fidelity, time-domain recordings of SPN-43 radar waveforms in the 3.5 GHz band.  For this purpose, a 60-second, complex-valued (i.e., I/Q) waveform was recorded roughly every ten minutes with a sample rate of 225~MS/s, and a corresponding spectrogram was computed.  The decision to retain a given waveform recording was made by comparing the spectrogram amplitudes to a threshold over the band of interest.  Although only a subset of the waveforms was retained for long-term storage, the entire set of spectrograms was saved.  

In total, 14,739 spectrograms were collected over the measurement campaign\footnote{The set of 3.5~GHz spectrograms used in this work is currently designated "For Official Use Only" (FOUO) by the U.S. Department of Defense, and is not publicly available.}.  Of these, approximately 58\% were acquired in San Diego and 42\% in Virginia Beach.  At each measurement site, data were collected with both an omni-directional antenna and a directional, cavity-backed spiral (CBS) antenna.  Roughly 45\% and 55\% of the spectrograms were acquired with the omni-directional and CBS antennas, respectively.  The spectrograms span a 200 MHz frequency range, typically 3465-3665 MHz, and a time-interval of one minute.  Each spectrogram has dimensions 134x1024, with 134 time-bins of duration 0.455 seconds and 1024 frequency-bins of length $225~\text{MHz}/1024 \approx 220~\text{kHz}$.  

The spectrograms were computed by applying a short-time Fourier transform (STFT) and then retaining the maximum amplitude in each frequency bin (i.e., a max-hold) over each 0.455 second time-epoch.  The window function for the discrete STFT was 1024 samples long, with the middle 800 points given a weight of one, and the left-most and right-most 112 points weighted with a cosine-squared taper.  The STFT was implemented with 112-sample overlap between consecutive time-segments.  Each spectrogram value is the maximum of $10^5$ time-averaged amplitudes, where the averaging duration is $4.55~\mu s$, because the STFT effectively averaged over a 1024 sample ($4.55~\mu s$) time window. 

When noted below, the spectrogram values were converted to power units (dBm) as follows.  Each max-hold spectrogram value was (i) divided by 1024, the STFT window length, (ii) divided by the (site-specific) front-end gain, (iii) multiplied by a measurement instrument calibration factor, (iv) squared and divided by $2\times 50$ (the 2 arises from the conversion between peak and root-mean-square (RMS) voltage for a narrowband signal, the 50 is for a 50 ohm load), and (v) converted to decibel-milliwatts (dBm) via the formula $10\log_{10}(P) + 30$, where $P$ is the power in Watts from step (iv).  For some calculations, the spectrogram values were further converted to dBm/MHz by subtracting $-6.2$~dBMHz, the effective noise bandwidth of the time-domain window for the STFT \cite[p.~32]{TN1954}.  

\begin{figure*}[!t]
		\centering
		\includegraphics[width=.32\textwidth]{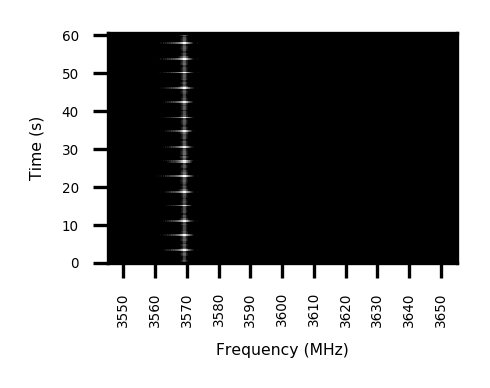}
		\includegraphics[width=.32\textwidth]{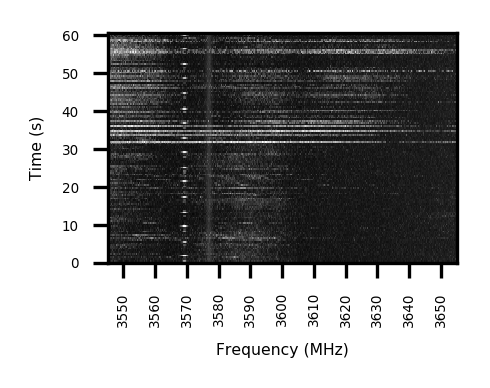}
		\includegraphics[width=.32\textwidth]{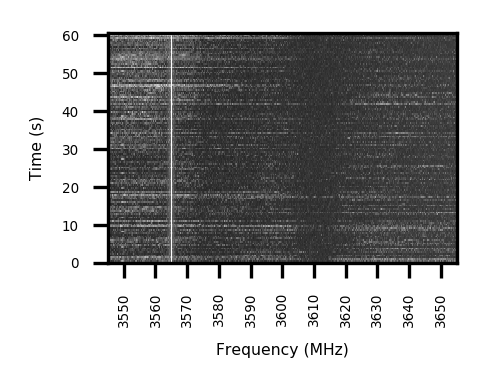}
		\caption{Example spectrogram captures, cropped to 3545-3655 MHz. (Left) Strong SPN-43 emissions near 3570~MHz; grayscale window [-90~-50]~dBm (Middle) Radar~3 OOBE coincident with SPN-43 emissions near 3570~MHz; grayscale window [-95~-75]~dBm (Right) Radar~3 OOBE coincident with weak SPN-43 emissions near 3550~MHz; grayscale window [-95~-75]~dBm.}
		\label{fig:spectrogram_examples}
	\end{figure*}

Figure~\ref{fig:spectrogram_examples} contains example spectrograms, cropped to the 3550-3650~MHz band of interest for ESC detection.  The original full-bandwidth (approx. 3465-3665 MHz) versions of these cropped spectrograms are shown in Figures~3.6 and 3.14 of \cite{TN1967}, and Figure~3.18 of \cite{TN1954}, for the \ifbool{double_column}{left}{top}, middle and \ifbool{double_column}{right}{bottom} spectrograms, respectively.  In each spectrogram, leakage from the local oscillator of the receiver is faintly visible as a vertical line at 3577~MHz (\ifbool{double_column}{Left}{Top} and Middle) and 3565~MHz (\ifbool{double_column}{Right}{Bottom}). The \ifbool{double_column}{left}{top} spectrogram contains a clear SPN-43 radar emission, located at approximately 3570~MHz.  Periodic radar sweeps are visible roughly every 4 seconds, corresponding to the SPN-43 antenna rotation period.  The middle and \ifbool{double_column}{right}{bottom} spectrograms of Figure~\ref{fig:spectrogram_examples} give examples of coincident Radar~3 OOBE and SPN-43.  In these images, Radar~3 OOBE are visible as horizontal streaks, and weak SPN-43 emissions are visible at 3570~MHz (Middle) and 3550~MHz (\ifbool{double_column}{Right}{Bottom}), respectively. 

The goal of this work was to create classifiers to identify SPN-43 presence in spectrograms.  In order to train and test classifiers, we needed a set of SPN-43 labeled data.  From the complete collection of 14,739 spectrograms, 4,491 were labeled by one of the co-authors for SPN-43 presence and Radar~3 OOBE.  Note that the human-applied labels are unverified, and based on subjective visual interpretation, as we did not have access to ship locations or assigned frequencies during or after the measurements.  Nearly 74\% (3,318) of the labeled spectrograms were selected for labeling because they correspond to captures that triggered retention of a recorded waveform.  Since this subset suffers from selection bias, an additional 1,173 spectrograms were randomly selected for human-labeling to provide a more diverse set of labeled cases.

\section{Classifiers}
	\label{sec:classifiers}
	This section gives implementation details for the thirteen classifier models that we evaluated for SPN-43 detection. Specifically, we compared two methods based on classical signal detection theory, three classical machine learning algorithms, and eight deep learning algorithms to the task of detecting SPN-43 in spectrograms.  Because we had a high degree of certainty about the data labels, we focused mainly on supervised learning methods.

For our classical signal-detection algorithms, we chose standard energy detection and sweep-integrated energy detection, which combines data integration with energy detection.  Further details can be found later in Section~\ref{sec:ed}.  The classical machine learning algorithms that were evaluated included Support Vector Machines (SVM), a K-Nearest Neighbor (KNN) classifiers, and Gaussian Mixture Models (GMM). Details are provided in Section~\ref{sec:ml}.

The deep learning methods included six well-known CNN architectures: VGG-16 and VGG-19 \cite{Simonyan2014}, ResNet-18 and ResNet-50 \cite{He2016}, the Inception-V1 network \cite{Szegedy2015}, and DenseNet-121 \cite{Huang2017}.  In addition, we implemented a CNN and an LSTM \cite{Hochreiter1997} of our own design.  For further details on the well-known architectures, see Section~\ref{sec:dl}. Details about our CNN and LSTM models are provided in Section~\ref{sec:cnn} and Section~\ref{sec:lstm}, respectively.  

The deep learning algorithms were implemented using the open-source TensorFlow\textsuperscript{TM} Python library running on an Nvidia\textsuperscript{\textregistered} DGX\textsuperscript{TM} workstation with four Tesla\textsuperscript{\textregistered} V100 Graphics Processing Unit (GPU) cards\footnote{Certain commercial software and hardware products are identified to fully specify our implementation. This does not imply endorsement by the National Institute of Standards and Technology or that the software and hardware are the best available for the purpose.}.  Details on standard deep learning layers and other elements, including their formulas, can be found in the textbook by Goodfellow et al. \cite{Goodfellow2016}.

Because the ESC detection task requires SPN-43 detection in each 10~MHz channel, the classifiers were first designed for the reduced task of detecting SPN-43 in a single 10~MHz channel.  For this purpose, spectrograms were divided into 10 MHz-wide channels centered at multiples of 10~MHz, e.g., 3550~MHz, 3560~MHz.  A single instance of each machine learning classifier was trained using a random sample of cases across all 10~MHz channels covering the 3550-3650 MHz band; further details on the training set are provided in Section~\ref{sec:training_set}.  Subsequently, to classify all 10~MHz channels over the 3550-3650~MHz band, copies of each classifier instance were applied in parallel, i.e., the same previously-trained classifier instance was applied to each channel.  In our preliminary investigations, we explored the possibility of using a single multi-channel detector rather than multiple single-channel detectors in parallel, but found that the latter improved detection performance.  Note that the SPN-43 classification results for the eleven channels covering 3550-3650 MHz were not fused for further decision-making. 
	
	\subsection{Classical Signal Detection Methods}
	\label{sec:ed}

From the family of classical signal detection methods, we evaluated two different approaches suitable for incoherent detection from spectrograms.  Note that techniques intended for coherent detection from in-phase and quadrature (I/Q) data were not possible with our data.  The first signal detection method was standard energy detection \cite{Urkowitz1967,Abdulsattar2012,Atapattu2014}.  The second method consisted of energy detection combined with data integration \cite[Sec. 1.4.6]{Richards2005}, a method commonly used in radar signal processing to increase signal-to-noise ratio (SNR).   Since this second method integrated the data over radar sweeps, we call it sweep-integrated energy detection.

Energy detection is a classical strategy based on the assumption that a signal of interest can be detected based on the total energy across a given time and frequency range. The total energy across the entire input is summed. If a given detection threshold by this summation is exceeded, the signal is decided to be present.

To improve the performance of energy detection, the whole 10~MHz channel was not used. Instead, only the 3 middle spectrogram columns (approximately 660 kHz) of each 10~MHz channel were aggregated for energy detection.  This range was chosen based on the results of an empirical evaluation.  Because SPN-43 can generally be expected to have carrier frequencies near multiples of 10~MHz, this modification excluded confounding emissions from the rest of the channel.

For sweep-integrated energy detection, we performed energy detection on specific parts of the input that would have highest SNR in the presence of SPN-43 and, therefore, hold the most information about signal presence.  Sweep-integrated energy detection is a form of data integration \cite[Sec. 1.4.6]{Richards2005}, a method commonly applied in radar signal processing.  First, a SPN-43 sweep template was generated, where the distance between sweep peaks was roughly $3.85$\,s \cite{TN1954, TN1967}.  This template consisted of a square wave in which the value was one for $0.455$\,s every $3.85$\,s and zero otherwise.  The point of highest cross-correlation was used to align the sweep template and the spectrogram.   Finally, the aligned template was applied as a mask to extract the portion of the input to use for standard energy detection, implemented as above. 

As detailed in \cite{TN1954,TN1967}, the spectrogram captures were collected with different front-ends at the two measurement sites.  To account for this fact, we applied site-dependent corrections to normalize the spectrograms to dBm units, as described in Section~\ref{sec:spectrograms}.  This normalization was only used for these two energy detection algorithms; the machine learning methods did not require any data normalization.  

Note that the two energy detection methods described above incorporate \emph{a priori} information.  First, both algorithms rely on the fact that SPN-43 typically has a carrier frequency that is a multiple of 10~MHz.  Second, the sweep-integrated energy detection method uses the fact that SPN-43 radar has a known sweep period.      
	
	\subsection{Classical Machine Learning Methods}
	\label{sec:ml}

For classical machine learning algorithms, we evaluated KNNs, SVMs, and GMMs \cite{Murphy2012,Zaki2014}. 

The KNN classification relies on similar data being spatially co-located within a chosen representational space.  When a new sample is classified, the most common label of the $k$-nearest labeled samples is assigned.  We evaluated the performance of all $k$ values for $k \in \{2, 5, 9, 12\}$.

The SVM classification attempts to maximize the distance between two classes and a parameterized hyperplane that separates the two classes, again, in some representational space.  A sample is assigned a label based on which side of the hyperplane it lies.  In addition to this standard type of SVM, called a linear SVM, distances can be measured using a kernel function which defines distances in alternate feature spaces. We evaluated the performance of linear SVMs, as well as SVMs with Radial Basis Function (RBF), polynomial, and sigmoid kernel functions \cite[Sec.~14.2]{Murphy2012}.

While GMMs are unsupervised, we examined them because it was one of the methods explored in \cite{Thilina2013}. A GMM fits a mixture of normal distributions with the expectation maximization algorithm. New samples are assigned the label with the highest posterior probability. Because the task was binary detection, the GMM was fit with a mixture of two normal distributions.

For each of the above algorithms, we evaluated the performance on the full channel and with two forms of data preprocessing. For the entire channel as input, the input was $134\times 46 = 6164$, because there were $134$ time steps and $46$ frequency bins per channel.

For the preprocessing, the goal was to reduce the dimensionality of the input.  In the first form of preprocessing, we applied energy aggregation across frequencies in the 10~MHz channel by summing the energy in each time step, which reduced our input vector length to $134$.  In the second form of preprocessing, we used prior information about SPN-43 for dimensionality reduction. Namely, because the carrier frequency for SPN-43 is typically a multiple of 10~MHz, the middle two frequency bins were extracted from the entire channel, resulting in an input size of $134\times 2 = 268$.

To signify which input we used for a given evaluation, we use subscripted model names. For example, KNN models are denoted as KNN$_{6164}$, KNN$_{134}$, and KNN$_{268}$, for the full, first preprocessed, and second preprocessed input forms, respectively.

	\subsection{Standard Deep Learning Classifiers}
	\label{sec:dl}
We evaluated the six standard deep learning CNNs listed above.  We chose these standard networks because most of them have been winners of a well-known competition using the ImageNet \cite{Imagenet09} benchmark dataset;  Liu et al. \cite{Liu2018} cited these networks as milestones in image classification.

VGG-16 and VGG-19\cite{Simonyan2014} were the first two models we evaluated. VGG-16 contains 13 convolutional layers, 3 fully connected layers, and 5 pooling operations. VGG-19 contains 16 convolutional layers, 3 fully connected layers, and 5 pooling operations. The VGG networks are powerful CNNs that, unlike more recent networks, only consist of convolutional layers, fully connected layers, and pooling operations.

Next, we evaluated ResNet-18 and ResNet-50 \cite{He2016}.  ResNet-18 consists of 17 convolutional layers, 2 pooling operations, and 1 fully connected layer. ResNet-50 consists of 49 convolutional layers, 2 pooling operations, and 1 fully connected layer. The ResNet networks introduce residual connections between blocks of convolutional layers which sum the outputs of a block with the input of the block. The benefit of these residual connections is the increased flow of information through backpropagation.

In addition, we evaluated Inception-V1 \cite{Szegedy2015}, which consists of 59 convolutional layers, 16 pooling operations, and 7 fully connected layers along with 2 local-response normalization operations.  This network was designed to handle different-sized features within the input feature space at each level of convolutional processing. As a result, it can handle a broad range of tasks with minimal tuning.

The final network we evaluated was DenseNet-121 \cite{Huang2017}, which contains 120 convolutional layers, 5 pooling operations, and 1 fully connected layers.  DenseNet has slightly worse performance on most benchmarks.  However, the number of parameters required to achieve similar performance compared to other standard networks is significantly smaller.  For each convolutional layer in a convolutional block, DenseNet concatenates the outputs and then downsamples using 1x1 convolutions.

See the papers cited above for further details.

	\subsection{Long Short-Term Memory Recurrent Neural Network}
	\label{sec:lstm}
Figure \ref{fig:flowcharts} (right) summarizes our LSTM architecture.  In order to effectively use the LSTM, we split the 10~MHz channel along the time axis to create sequential slices of approximately 0.455 seconds.  Each of the time-slices was fed into the LSTM cell one at a time along with the previous output of the LSTM cell.  This is known as a residual connection.  Motivations for using residual connections in LSTMs include protection against the vanishing gradient problem in backpropagation \cite{Kim2017} as well as greater network expressivity \cite{He2016}.  Dropout was used between LSTM cells with a probability of 50\%.  

After all of the time slices were fed into the LSTM, the output of the last cell was passed on to a fully-connected layer with 50 neurons.  Next, a bias was added to the output of the 50-neuron fully-connected layer and a ReLU was applied. The output of the ReLU was then passed to a fully-connected layer of size 1 and a bias was added.  Lastly, a sigmoid activation was applied to the output to generate the prediction, a continuous-valued number between zero and one.  The LSTM was trained using stochastic gradient descent with cross-entropy loss; further details on training are given in Section~\ref{sec:training}.   

	\subsection{Convolutional Neural Network}
	\label{sec:cnn}
\begin{figure}
		\begin{subfigure}{.5\columnwidth}
			\centering
			\includegraphics[height=.6\textheight]{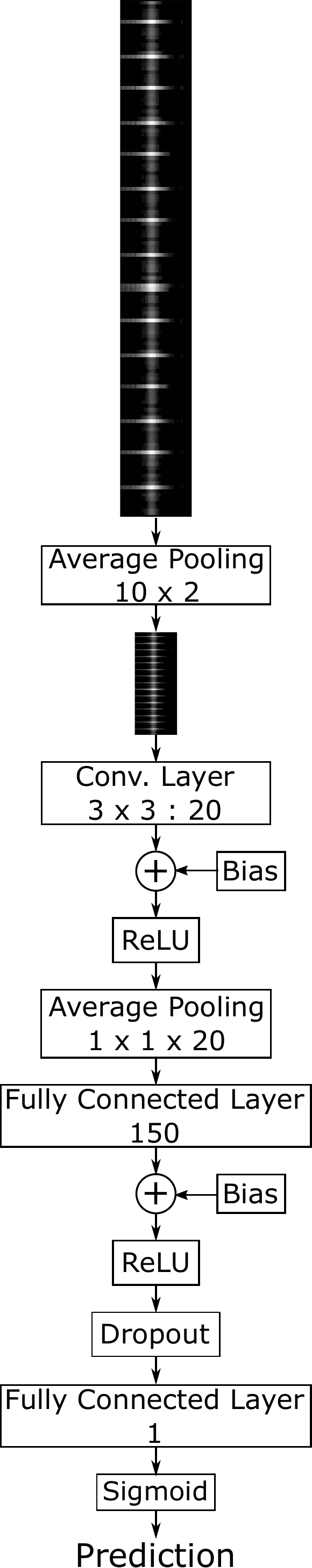}
		\end{subfigure}%
		\begin{subfigure}{.5\columnwidth}
			\centering
			\includegraphics[height=.6\textheight]{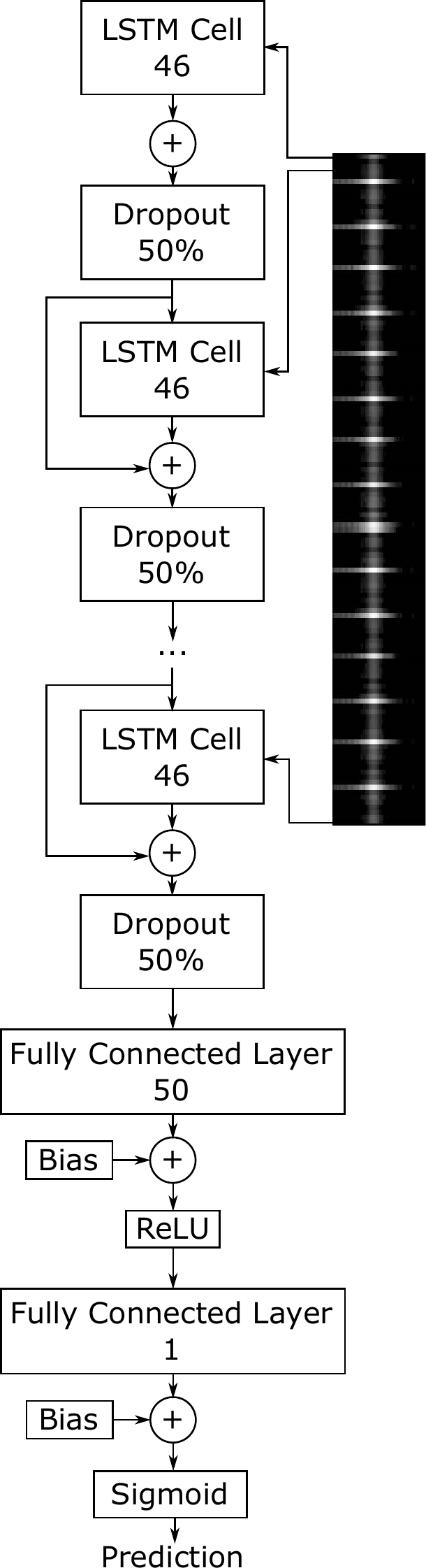}
		\end{subfigure}
		\caption{Flowcharts for the deep learning implementations.  (Left) The CNN-3 architecture. (Right) The LSTM architecture.}
		\label{fig:flowcharts}
	\end{figure}

Figure \ref{fig:flowcharts} (left) summarizes our CNN architecture, henceforth referred to as CNN-3 for its three layers, as explained below.  First, the 10~MHz channel was passed through an average pooling operation with a window size of 10x2, resulting in a down-sampled spectrogram with time and frequency dimensions reduced by factors of 10 and 2, respectively.  The down-sampled spectrograms were then passed to a convolutional layer with 20 filters of size 3x3 and stride 1x1. Zero-padding was not used.  Subsequently, a bias (i.e., constant) was added to the filter activations and a rectifier linear unit (ReLU) was applied to the resulting output. The output of the ReLU step consisted of 20 activation maps for each of the convolutional layer's filters.

Next, the activation maps were averaged together to create a single averaged-activation map using an operation we call channel-average pooling.  Specifically, this operation averages co-located values across all channels and averages them to output a single channel. Equivalently, this operation can be described as a single-filter 1x1 convolutional layer \cite{Lin2013} in which all weights are equal and sum to 1. To our knowledge, this averaging step has not been suggested previously in the CNN literature although it does resemble other channel pooling methods \cite{Huang2015}.

Note that above channel-average pooling operation is distinct from conventional average pooling, which takes localized areas from within a single channel, where a channel is a depth slice of the output from a convolutional layer, and outputs the same number of channels with average values for these areas.  We found that using the averaged activations rather than the individual activation maps showed empirical improvements in accuracy.

The output of the channel-average pooling operation was passed into a fully-connected layer containing 150 neurons. A bias was added to the output of this fully-connected layer and a ReLU was applied. The output of the ReLU was then fed through a dropout step, with a dropout probability of 50\%.  Subsequently, the output from the dropout step was fed into another fully-connected layer containing a single neuron followed by a bias.  Finally, the biased output was passed through a sigmoid activation function to produce the prediction, a continuous-valued number between zero and one.  CNN-3 was trained using stochastic gradient descent with cross-entropy loss; further details on training are given in Section~\ref{sec:training}.

\section{Classifier Performance Assessment}
	As noted in Section~\ref{sec:spectrograms}, a set of 4,491 spectrograms were labeled for SPN-43 presence and Radar~3 OOBE.  This collection of labeled data was partitioned into two disjoint sets: one for training and one for testing.  In this section, after explaining how the two sets were selected, we present performance results for each classifier.  Since ESC detection will only be required for the 3550-3650 MHz portion of the 3.5 GHz band, the training and testing sets described below were limited to the eleven 10~MHz channels covering the range 3545-3655, with each channel centered on multiples of 10~MHz.

	\subsection{Test Set Composition}
		The sample of labeled spectrogram data was potentially biased in two respects.  First, because the data collection was observational, at only two geographic locations for two months each, the respective proportions of different emission types did not necessarily reflect those in the whole population of possible field measurements, i.e., the distribution of 3.5~GHz emissions at all coastal locations under all conditions.  Second, as mentioned in Section~\ref{sec:spectrograms}, nearly 74\% of the labeled spectrograms were selected for labeling because they corresponded to captures that triggered retention of a recorded waveform.  Consequently, the set of labeled data suffered from a selection bias that resulted in a disproportionate number of labeled spectrograms with high-amplitude emissions.  

Due to the above potential biases in the labeled data set, and due to the need for sufficient testing of important sub-groups, a stratified sampling approach was utilized to construct the test set.  Specifically, a test set was randomly selected from the set of labeled data with approximately equal proportions of spectrograms across emission categories (SPN-43, Radar~3 OOBE, Both SPN-43 and Radar~3 OOBE, Neither), measurement locations (Virginia Beach and San Diego), and antenna type (Omni-directional and CBS).  In addition, the maximum number of spectrograms containing multiple SPN-43 emissions were included.  Table~\ref{test_set_composition} shows the proportions for each category in the most general test set, denoted Test Set A.  Note that the proportions are not exactly equal because the random test set generation program had to satisfy a hierarchy of preferences that did not typically lead to a perfect solution.  Also, observe that roughly 50\% of the cases contained SPN-43 and 50\% of the cases did not contain SPN-43.  To assess classifier performance without the presence of Radar~3 OOBE, a subset of Test Set A, called Test Set B, was used; see Table~\ref{test_set_composition}.  

The data stratification used for the test set was selected to ensure that the full gamut of test cases were represented, including variations in measurements due to channel effects, receiver reference level, antenna type, measurement location, etc.  Our aim was to carry out a rigorous evaluation of models by including an adequate representation of all cases that are likely to be observed in the field. 
 
		\begin{table*}[tb]
			\centering
			\small
			\begin{tabular}{|c | c | c || c | c | c | c || c | c || c | c |}
				\hline
				\multicolumn{3}{|c||}{ } & \multicolumn{4}{|c||}{Emissions}  & \multicolumn{2}{|c||}{Location}& \multicolumn{2}{|c|}{Antenna} \\ \hline
				 Set & Total & Multi SPN-43 & SPN-43 & R3-OOBE & Both & Neither & VB & SD & Omni & CBS \\ \hline
				 A & 509 & 109 & 24.56\% & 24.36\% & 26.72\% & 24.36\% & 51.28\% & 48.72\% & 48.92\% & 51.08\% \\
				 B & 249 & 40 & 50.20\% & 0\% & 0\% & 49.80\% & 50.20\% & 49.80\% & 50.20\% & 49.80\% \\
				 \hline
			\end{tabular}
			\caption{Test Set Compositions.  Above, ``R3-OOBE'' stands for Radar~3 OOBE.  Test Set B is a proper subset of Test Set A that does not contain cases with Radar~3 OOBE.}
			\label{test_set_composition}
		\end{table*}

	\subsection{Training}
		\label{sec:training}
		\label{sec:training_set}

As stated in Section~\ref{sec:classifiers}, the machine learning classifiers were first trained on 10~MHz channels for single-channel detection and then these already trained single-channel detection instances are connected in parallel for multichannel detection over the full spectrogram.  The training set consisted of 10~MHz channels randomly extracted from the collection of labeled spectrograms not in Test Set A, which included cases collected at both measurement locations, with both antenna types, and all receiver reference levels.  A total of 4,285 channels were randomly selected for training, where half contained SPN-43 and half did not.

All deep learning methods used uniform Xavier initializations \cite{Glorot2010} for convolutional layers; truncated mean initializations with means of zero, standard deviations of one, and truncation at two standard deviations above and below the mean for fully connected layers; and zero initializations for bias layers.  Both Adagrad \cite{Duchi2011} and Adam \cite{Kingma2014} optimizers were compared during training with no empirical advantage between the two. The learning rate was fixed at $0.0001$ for all algorithms and the cross-entropy loss function was used. 

\begin{figure}[tb]
			\centering
			\includegraphics[width=4.3cm]{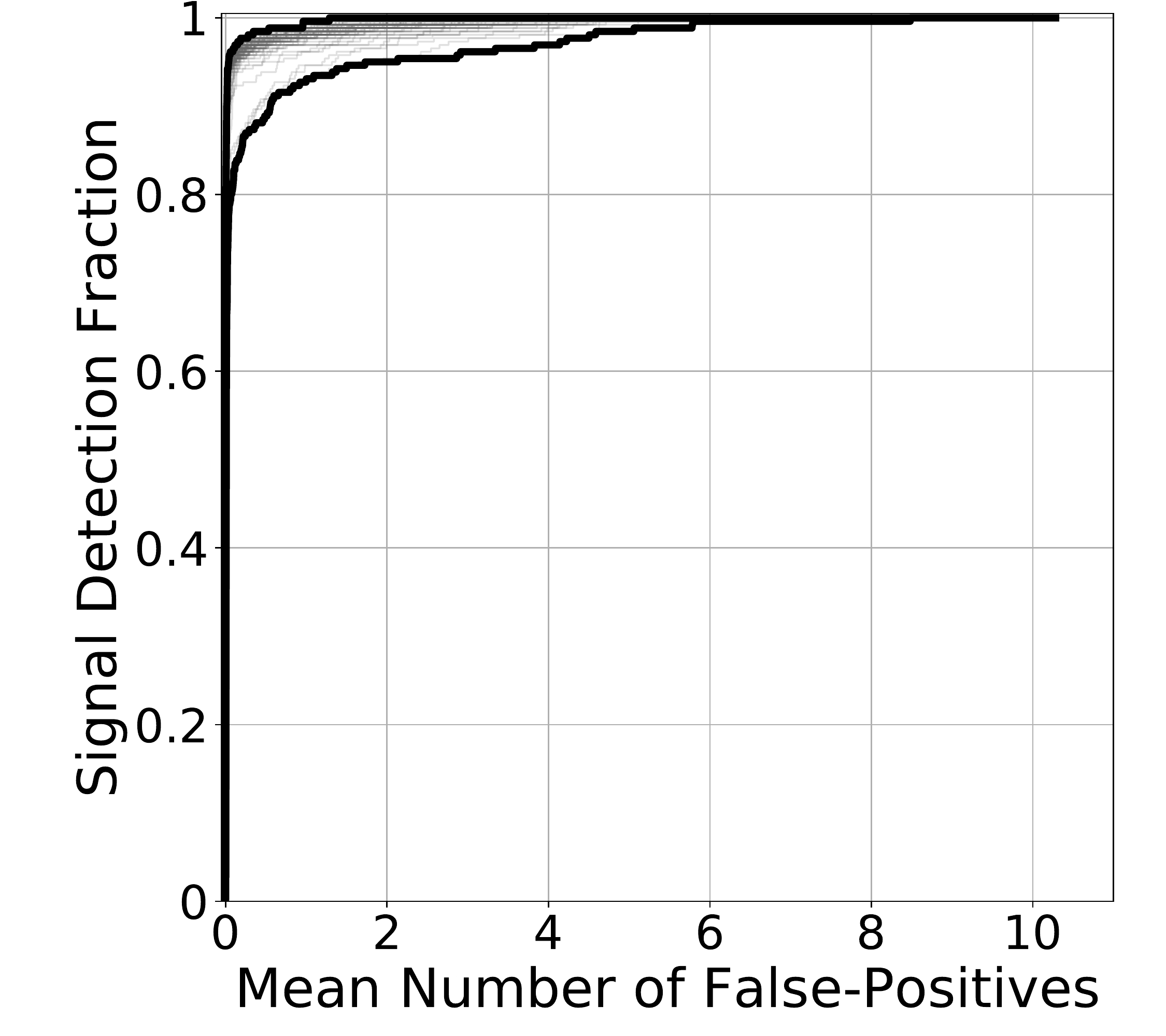}
			\includegraphics[width=4.3cm]{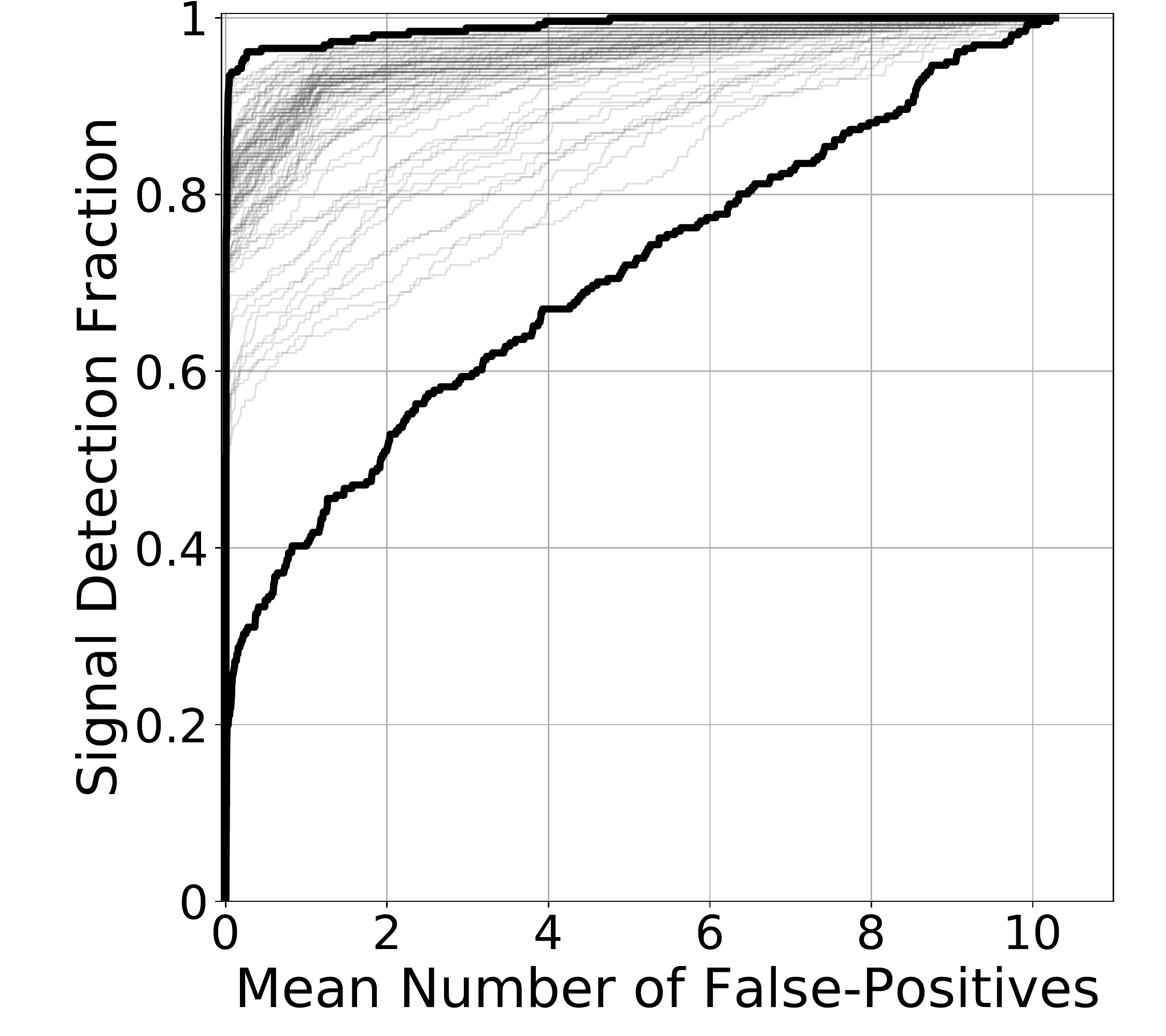} 
			\caption{FROC curves summarizing multichannel detection performance on Test Set A over 100 different training initializations for CNN-3 (Left) and LSTM (Right).  The minimum and maximum bounds over all initializations are shown in bold.}
			\label{fig:init_variance}
		\end{figure}

To assess the sensitivity of CNN-3 and the LSTM training to weight initialization, we trained both algorithms 100 times with different, randomly-generated initializations.  For this test, we used the Adagrad optimizer.  We used cross-entropy for the loss function.  Each training instance was run using the same training set for 1,000 epochs (an epoch consists of one pass through all training examples).  Figure~\ref{fig:init_variance} summarizes the performance of CNN-3 and the LSTM on Test Set A over the 100 training initializations.  In this figure, multichannel detection performance for each training initialization is summarized with an empirical free-response receiver operating characteristic (FROC) curve, shown in light gray.  Appendix~\ref{ROCappendix} reviews FROC curves, which can be used to summarize multichannel detection performance.  The minimum and maximum bounds over all 100 FROC curves are shown in bold.  Note that these bounds are not necessarily the same as any individual FROC curve.      

From the plots in Figure~\ref{fig:init_variance}, we can draw three conclusions.  First, the CNN-3 distribution is much tighter than the LSTM distribution.  Second, the best LSTM and CNN-3 instances performed similarly.  Last, these plots emphasize the necessity of testing classifier performance over multiple training initializations before settling on a particular set of weights.  The results presented below were generated using the CNN-3 and LSTM instance with the largest area under the FROC curve (FROC-AUC). 

	\subsection{Performance Evaluation}
		\label{sec:perf_eval}
		Single-channel and multichannel detection performance was assessed using receiver operating characteristic (ROC) and FROC curves, respectively.  See Appendix~\ref{ROCappendix} for an introduction to ROC and FROC curves.  In the single-channel ROC evaluations, the eleven 10~MHz channels covering 3550-3650~MHz in each spectrogram were tested, and the results were aggregated on a per-channel basis.  On the other hand, the multichannel FROC evaluations holistically assessed detection performance over the entire 3550-3650~MHz frequency range by aggregating detection results on a per-spectrogram basis.

Single-channel performance of all thirteen algorithms is evaluated. For conciseness, however, the multichannel performance is only evaluated for the top-performing algorithm of each class for Test Set~A.  Single-channel detection performance is summarized by estimates of the area under the ROC curve (ROC-AUC), where a higher number indicates better performance.  Only the top-performing SVM, KNN, and GMM models are shown in the single-channel analysis included here.  Specifically, the top-performing models were the linear SVM$_{6164}$, KNN$_{6164}$ with $k=9$, and GMM$_{268}$.

Table~\ref{table:AUC} gives estimates of the ROC-AUC on Test Set~A and B. Each AUC point estimate was estimated nonparametrically by computing the area under the empirical ROC curve.  This is mathematically equivalent to the normalized Mann-Whitney U statistic \cite{Pepe2003}.  Ninety-five percent confidence intervals were estimated using the nonparametric method of DeLong et al. \cite{DeLong1988} together with the logit transformation method recommended by Pepe \cite[p.~107]{Pepe2003}.  The top-performing algorithms for each algorithm category are in bold.  Note that because the classifiers were applied to the same test set, the confidence intervals are correlated.

Table~\ref{table:FAUC} gives estimates of the FROC-AUC on Test Set~A and B.  In this table, FROC-AUC was normalized to make it a number between 0 and 1; the normalization factors for test set A and B were 10.28 and 10.36, respectively.  See Appendix~\ref{ROCappendix} for a discussion of our rationale for FROC-AUC normalization.  The AUC point estimates were estimated nonparametrically by computing the area under the empirical FROC curves.  Ninety-five percent confidence intervals for FROC-AUC were estimated using the percentile bootstrap method \cite{Efron1993}, where the bootstrapping was stratified to maintain the proportions in Table~\ref{test_set_composition}.  Again, because the classifiers were applied to the same test sets, the confidence intervals are correlated.

\begin{figure}[p]
			\centering
			\includegraphics[width=4.3cm]{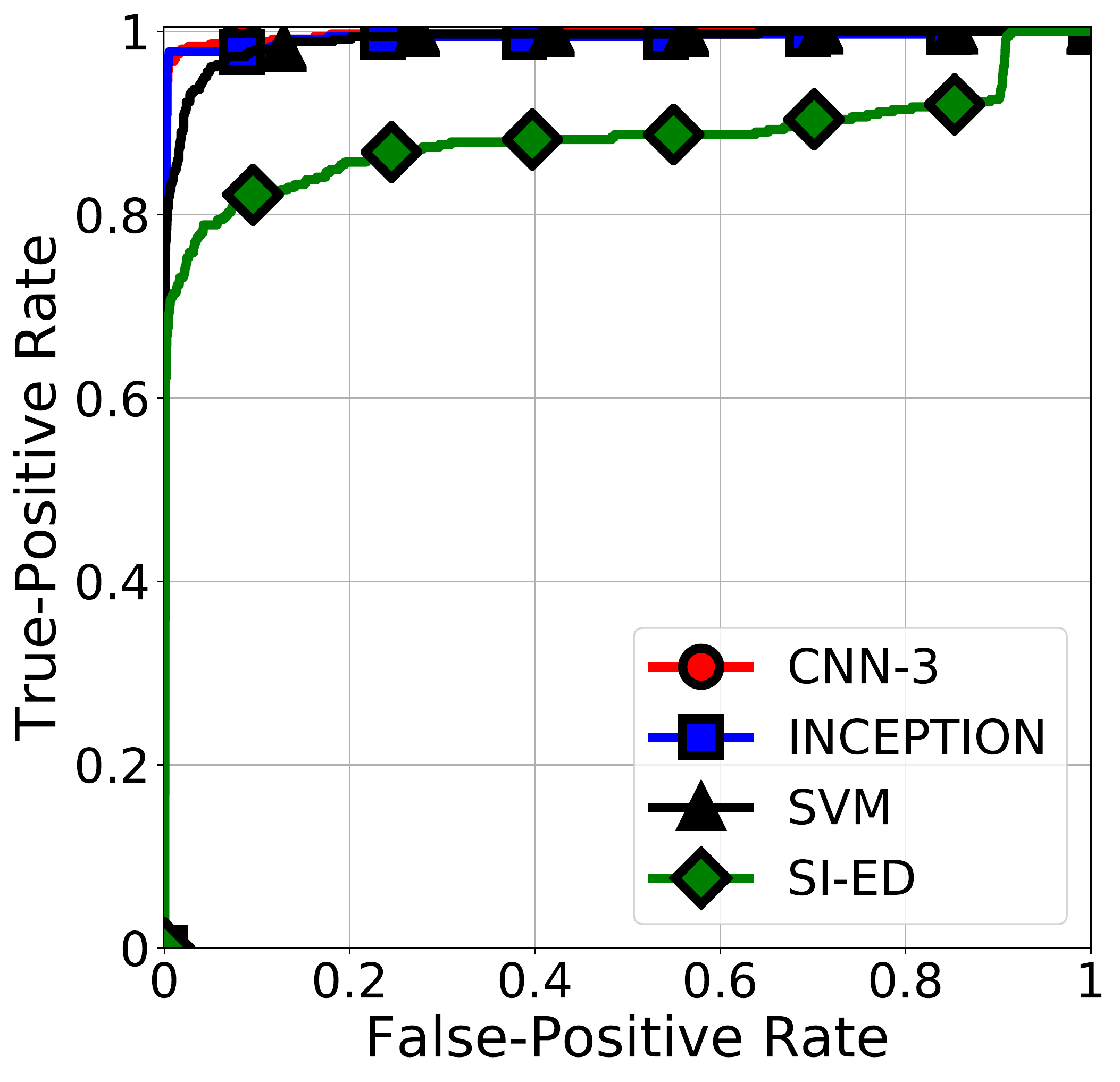} 
			\includegraphics[width=4.3cm]{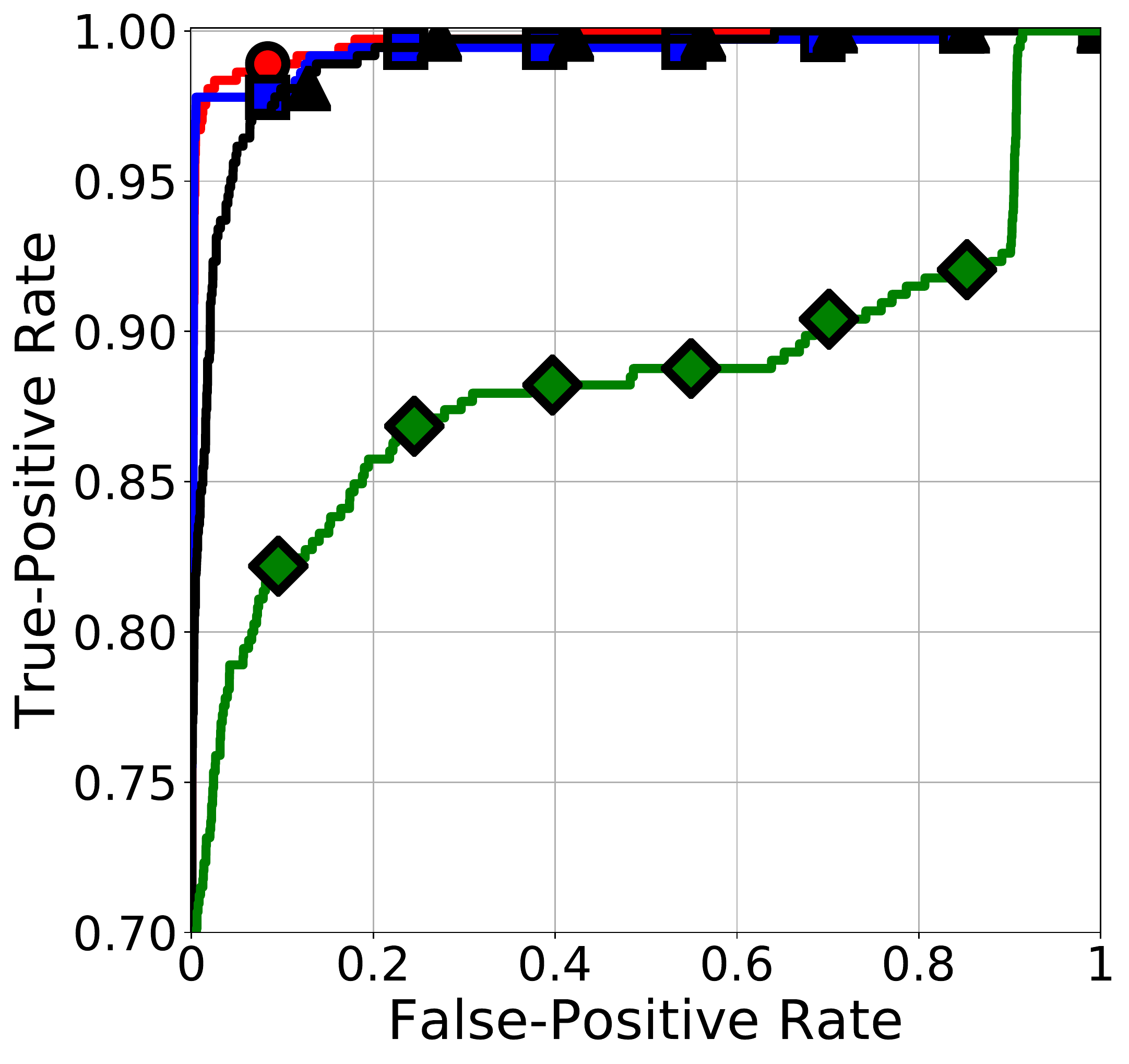} 
			\caption{Test Set A ROC results.  Left: Full ROC curves for single channel detection. Right: Y-axis zoom of plot on left.}
			\label{fig:setAroc}
		\end{figure}

From the ROC-AUC results, we see that all machine learning methods decidedly outperformed sweep-integrated energy detection.  However, there was not a statistically significant difference between CNN-3, Inception-v1, and SVM$_{6164}$.  The FROC-AUC results yield a similar conclusion.  Figures~\ref{fig:setAroc} and \ref{fig:setAfroc} show empirical ROC and FROC curves for CNN-3, Inception-v1, the linear SVM, and sweep-integrated energy detection (SI-ED) on Test Set~A.  These plots support the AUC conclusions, but provide additional insight into differences in performance across false positive rates.  

\begin{figure}[p]
			\centering
			\includegraphics[width=4.3cm]{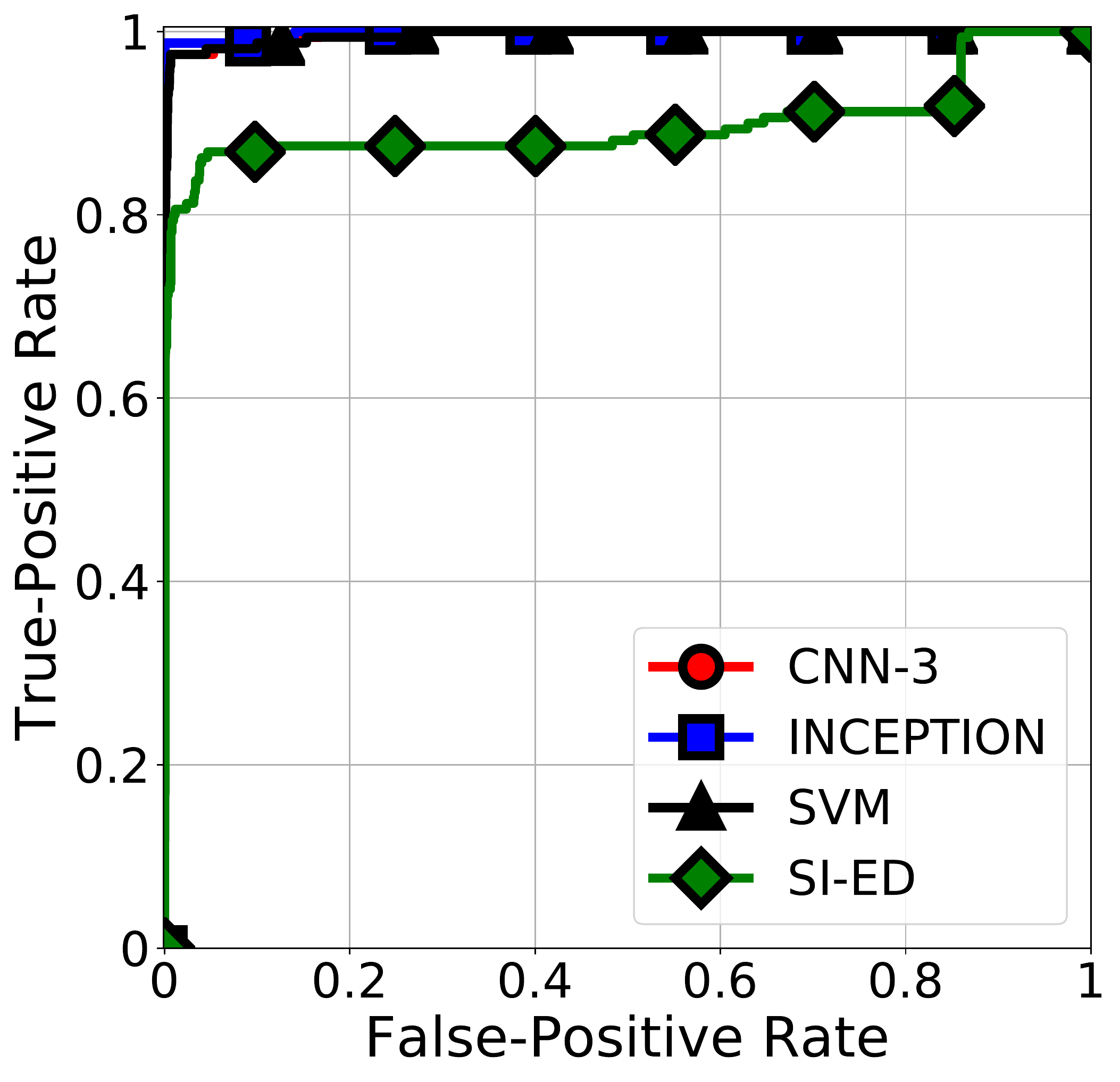} 
			\includegraphics[width=4.3cm]{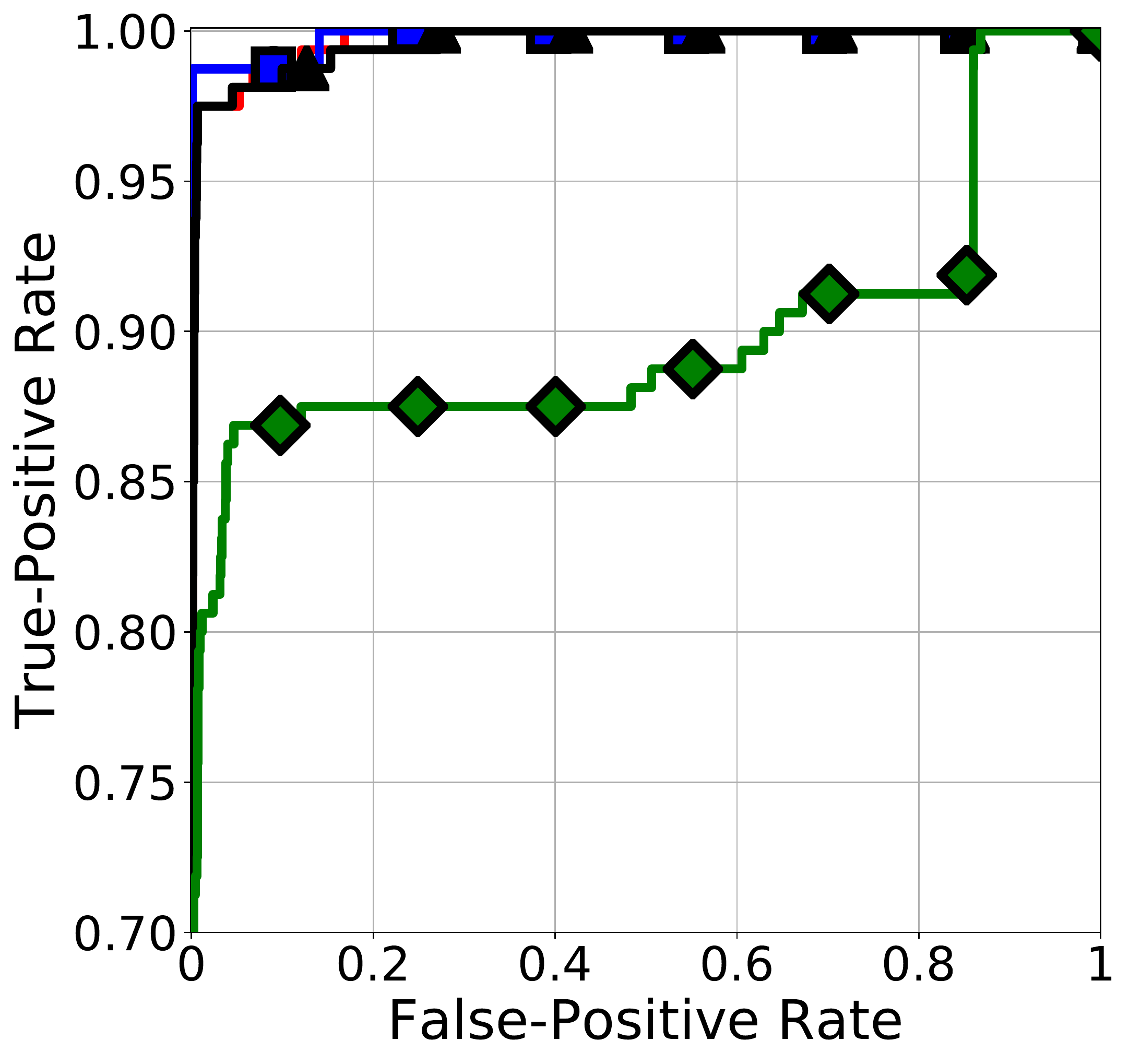} 
			\caption{Test Set B ROC results.  Left: Full ROC curves for single channel detection. Right: Y-axis zoom of plot on left.}
			\label{fig:setBroc}
		\end{figure}

\begin{figure}[p]
			\centering
			\includegraphics[width=4.3cm]{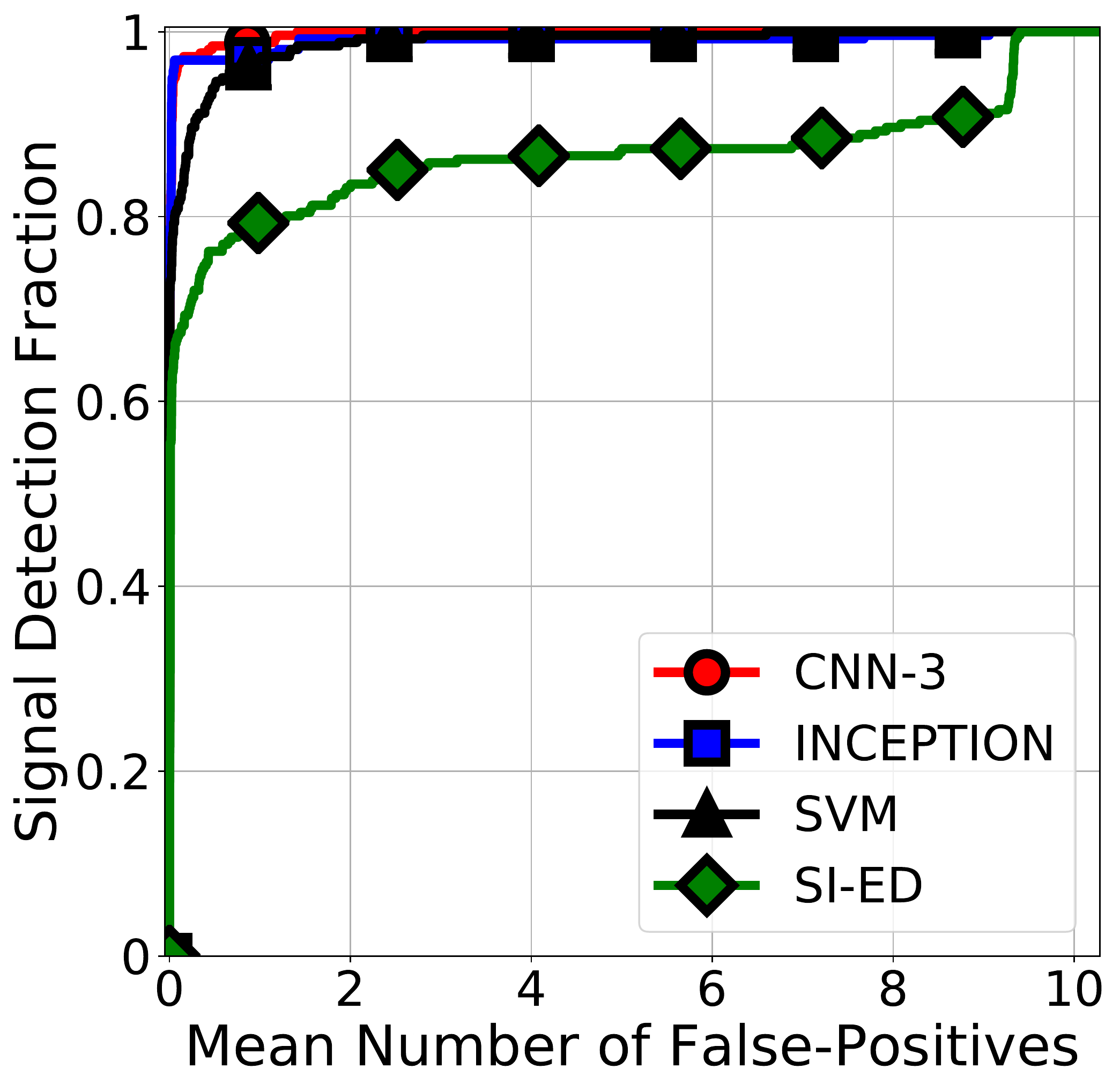} 
			\includegraphics[width=4.3cm]{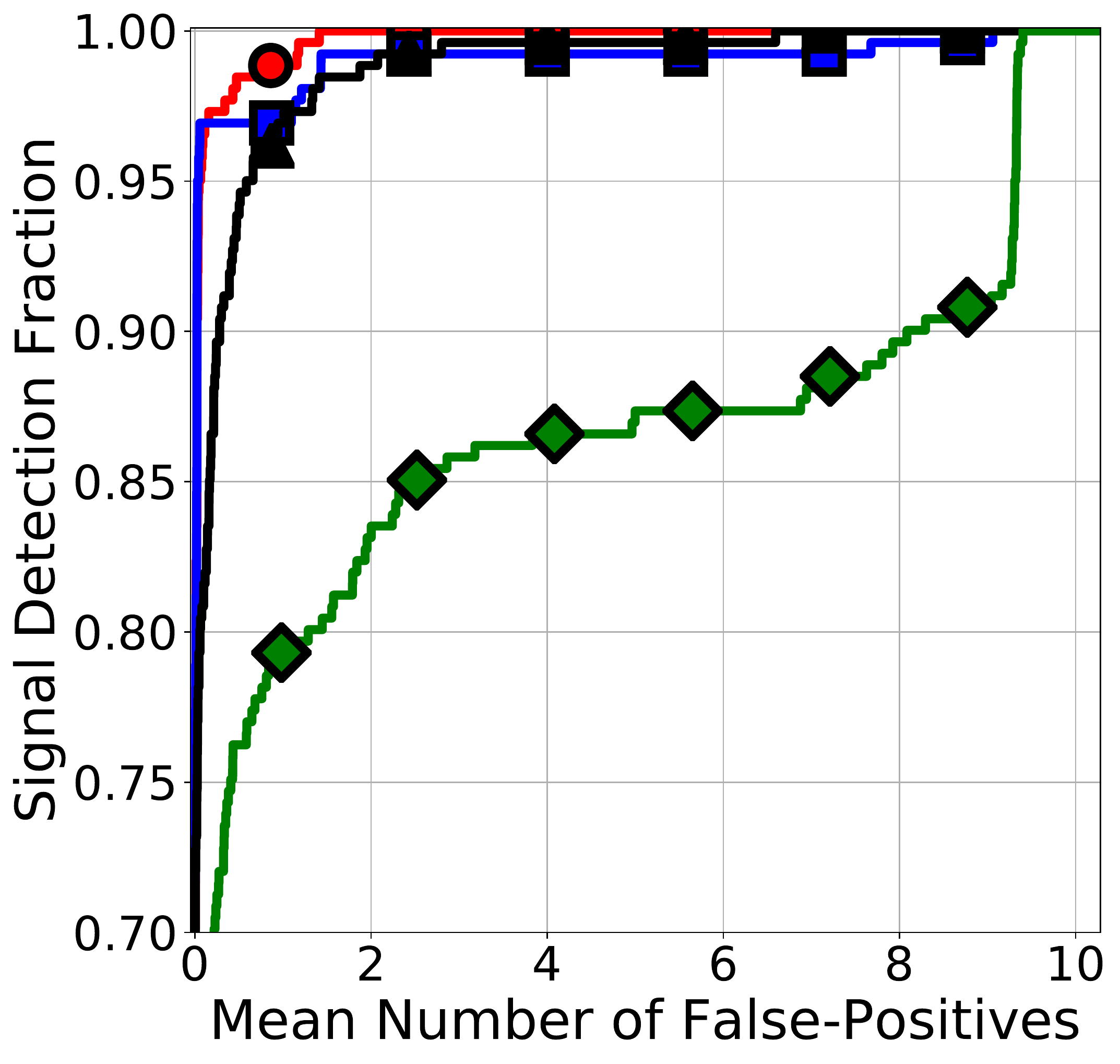} 
			\caption{Test Set A FROC results.  Left: Full FROC curves for multichannel detection. Right: Y-axis zoom of plot on left.}
			\label{fig:setAfroc}
		\end{figure}

\begin{figure}[p]
			\centering
			\includegraphics[width=4.3cm]{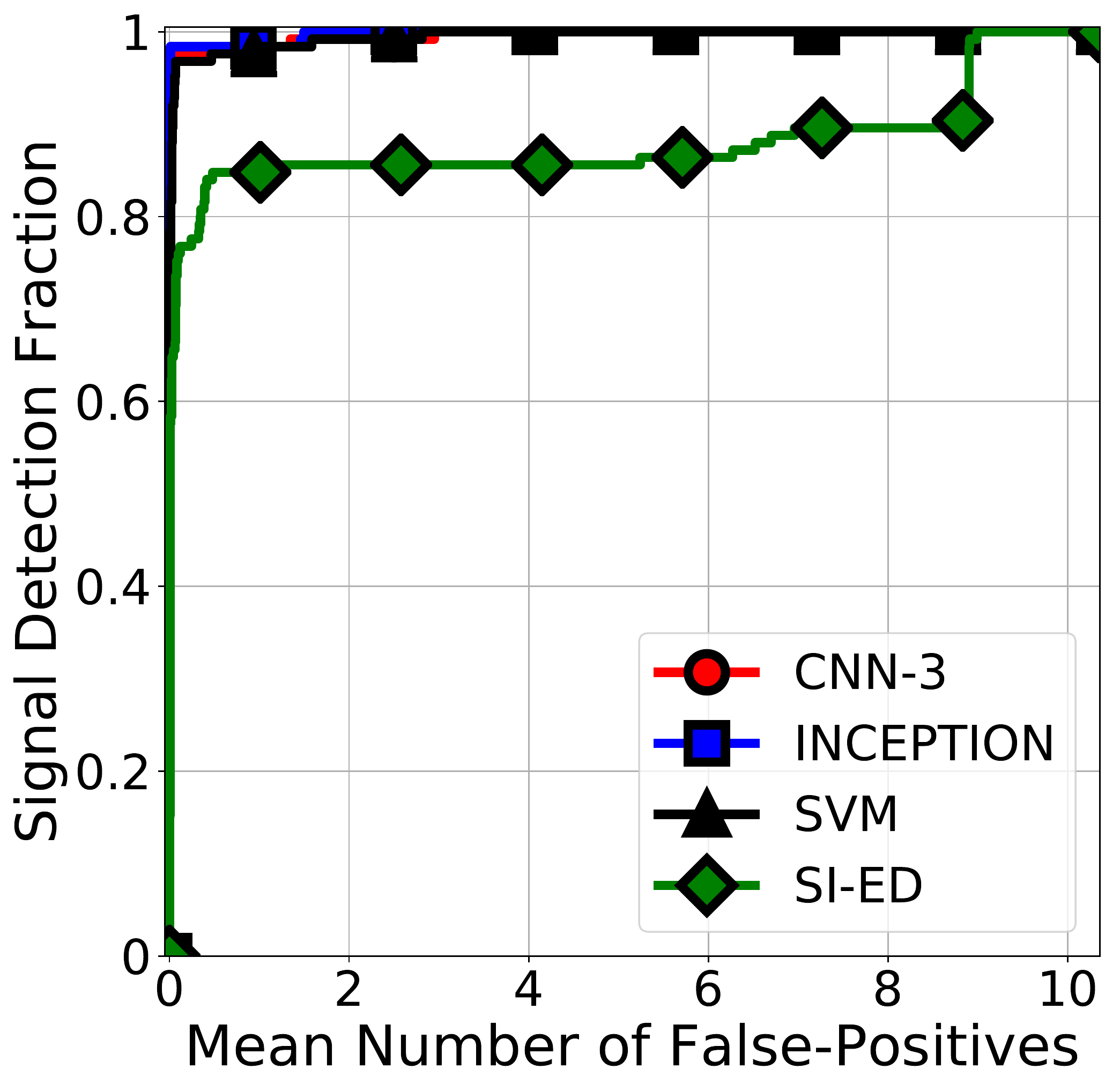} 
			\includegraphics[width=4.3cm]{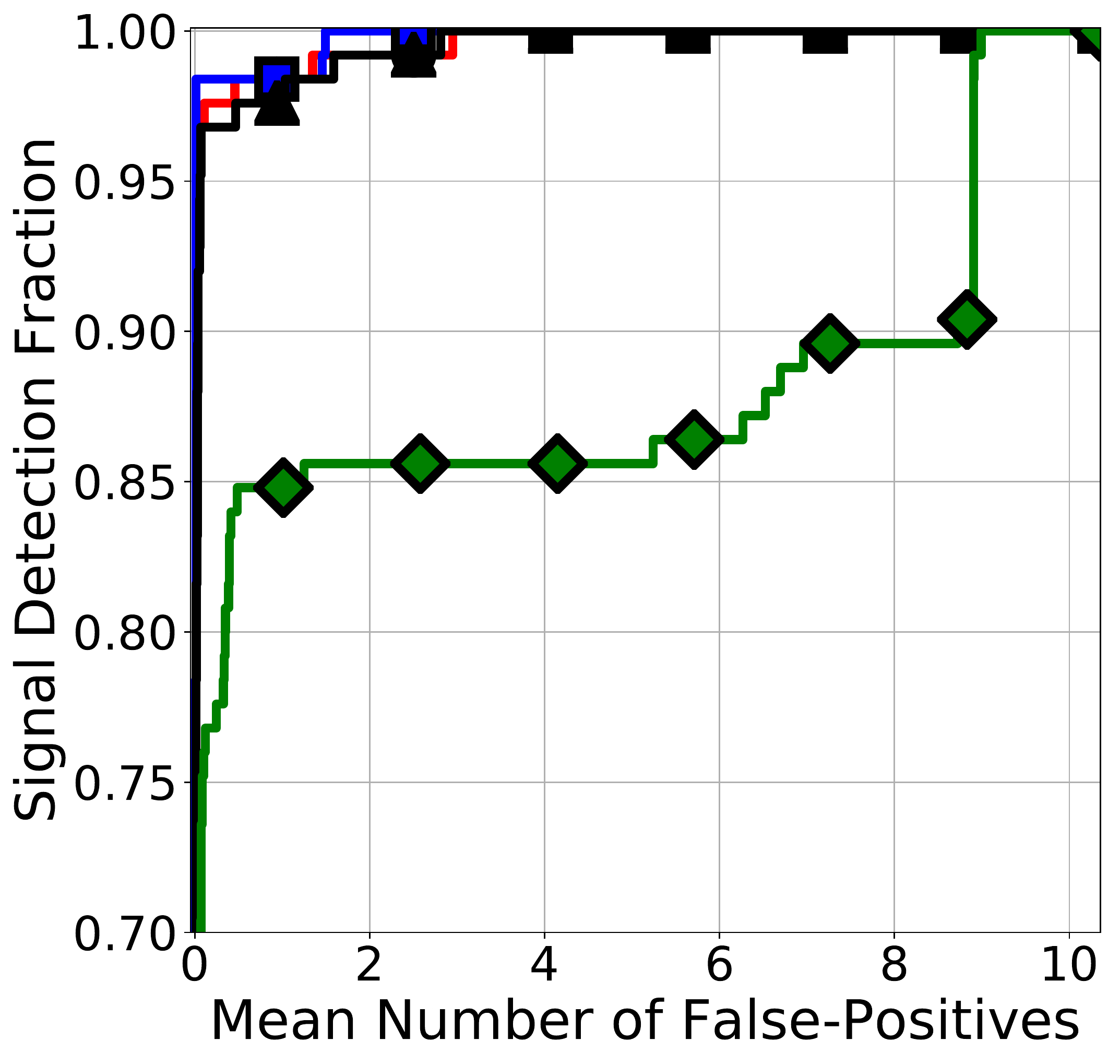} 
			\caption{Test Set B FROC results.  Left: Full FROC curves for multichannel detection. Right: Y-axis zoom of plot on left.}
			\label{fig:setBfroc}
		\end{figure}

\begin{figure}[htb]
			\centering
			\includegraphics[width=4.3cm]{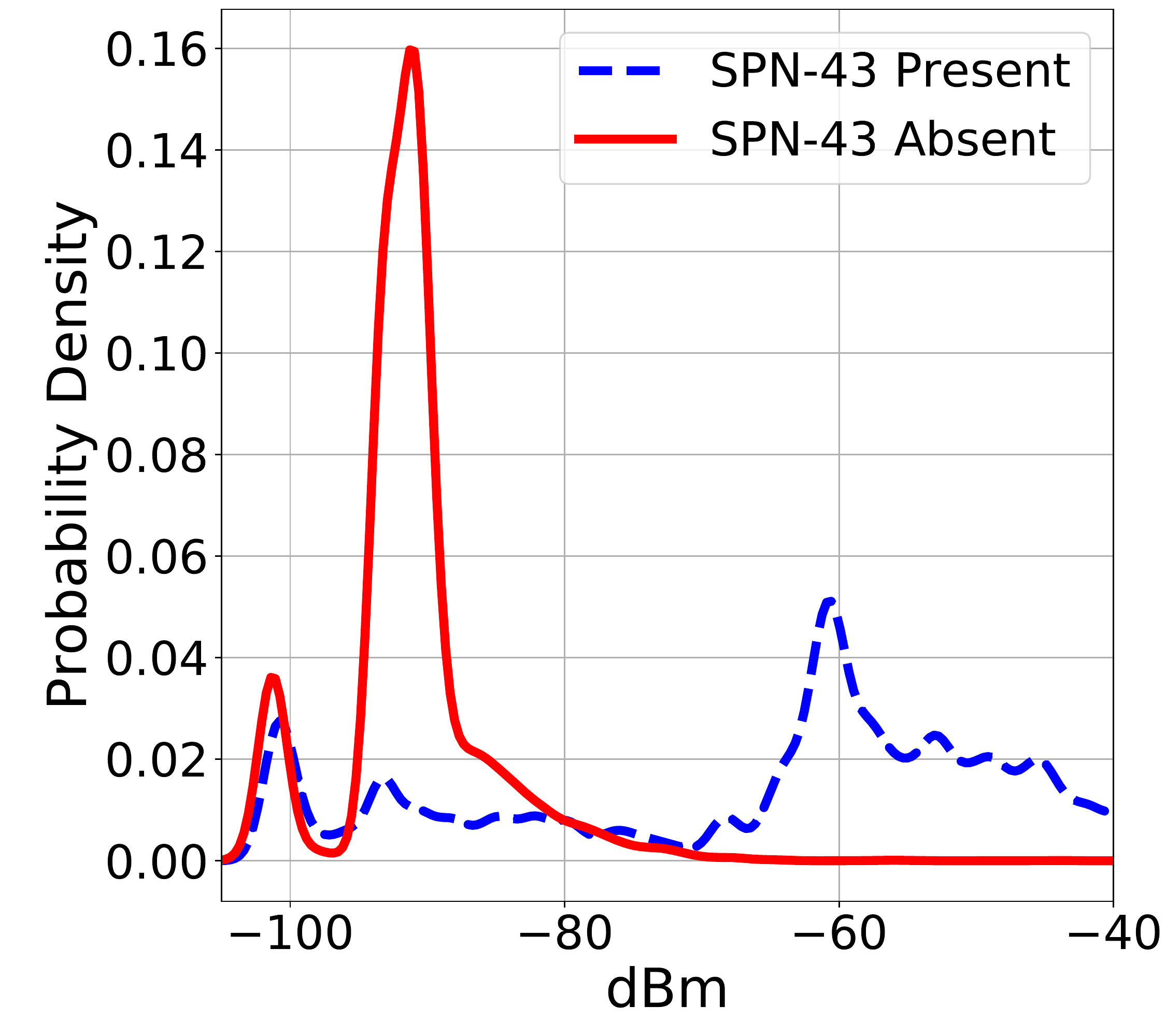} 
			\includegraphics[width=4.3cm]{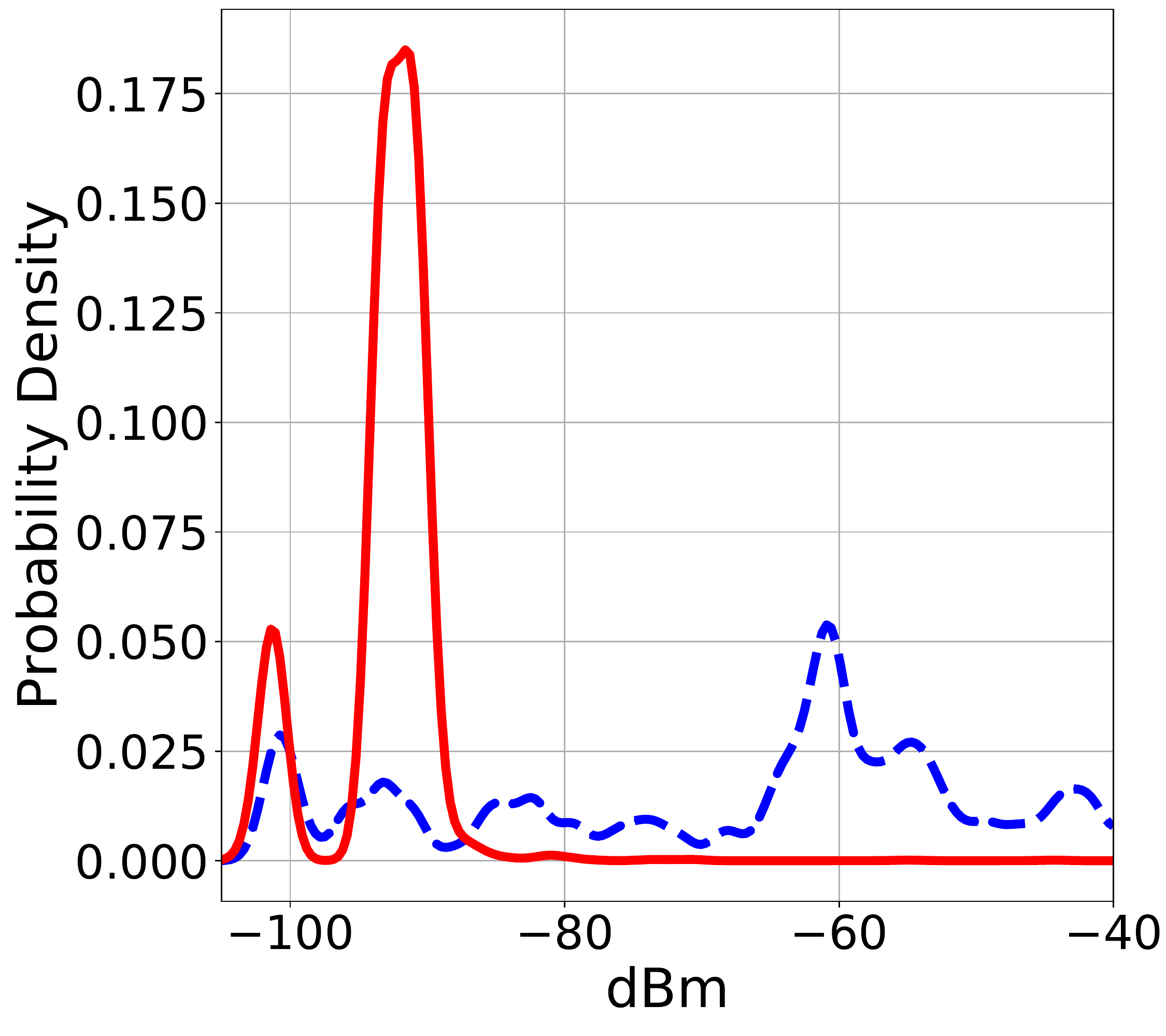} 
			\caption{Estimated probability density functions for power in the frequency bins used by energy detector.  Left: Test Set A. Right: Test Set B.}
			\label{fig:test_set_pdfs}
		\end{figure}

The detection performance for Test Set~B, a proper subset of Test Set~A that excluded Radar~3 OOBE, is summarized in Figure~\ref{fig:setBroc} and Figure~\ref{fig:setBfroc} and with the respective columns in Table \ref{table:AUC} and \ref{table:FAUC}.  Comparing these results to those for Test Set~A, we see that the removal of Radar~3 OOBE only yielded a slight improvement in sweep-integrated energy detection for low false-positive rates.  To elucidate this finding, Figure~\ref{fig:test_set_pdfs} shows estimated probability density functions (PDFs) for the power in each 660~kHz-wide channel used by the sweep-integrated energy detector for SPN-43-absent and SPN-43-present cases.  Each PDF was estimated using the kernel density estimation method with a Gaussian kernel and a bandwidth of one.  The peaks in the SPN-43-absent distributions correspond to the receiver noise floor, which varied with measurement location and receiver reference level \cite{TN1954,TN1967,TN2016}.  From these plots, we see that the SPN-43-absent PDF for Test Set~A has a fatter tail between -85 dBm and -70 dBm than for Test Set~B, which is consistent with Radar~3 OOBE being present in set A.  However, there is only a very slight shift in the peaks of each distribution for sets A and B. The very small differences between the distributions for sets A and B help to explain the small improvement in sweep-integrated energy detection performance on Test Set~B.

		\begin{table}[htb]
			\centering
			\begin{tabular}{|r | c | c |}
				\hline
				Classifier & ROC-AUC (Set A) & ROC-AUC (Set B) \\ \hline
				CNN-3 & \textbf{.997,~[.993, .998]} & \textbf{.997,~[.992, .999]} \\
				LSTM & .988,~[.978, .994] & .994,~[.987, .998] \\ \hline
				Inception-v1 & \textbf{.994,~[.985, .997]} & \textbf{.998,~[.993, .999]} \\
				ResNet-18 & .913,~[.889, .932] & .891,~[.856, .918] \\
				DenseNet-121 & .838,~[.807, .866] & .782,~[.727, .829] \\
				ResNet-50 & .622,~[.576, .665] & .623,~[.558, .684] \\
				VGG-16 & .500,~[.500, .500] & .500,~[.500, .500] \\ 
				VGG-19 & .500,~[.500, .500] & .500,~[.500, .500] \\ \hline
				SVM$_{6164}$ & \textbf{.991,~[.985, .994]} & \textbf{.996,~[.989, .998]} \\
				KNN$_{6164}$ & .965,~[.949, .976] & .981,~[.959, .991] \\
				GMM$_{268}$ & .868,~[.845, .888] & .884,~[.847, .913] \\ \hline
				SI-ED & \textbf{.884,~[.854, .910]} & \textbf{.899,~[.850, .933]} \\
				ED & .854,~[.819, .883] & .852,~[.796, .895] \\
				\hline
			\end{tabular}
			\caption{ROC-AUC estimates for Test Set A and B.  Each entry shows a point estimate and a 95\% confidence interval.}
			\label{table:AUC}
		\end{table}
	
		\begin{table}[htb]
			\centering
			\begin{tabular}{|r | c | c |}
				\hline
				Classifier & FROC-AUC (Set A) & FROC-AUC (Set B) \\ \hline
				CNN-3 & \textbf{.997,~[.994, .998]} & .993,~[.984, .997] \\
				Inception-v1 & .990,~[.980, .998] & \textbf{.997,~[.993, 1.00]} \\
				SVM$_{6164}$ & .988,~[.980, .993] & .994,~[.988, .999] \\
				SI-ED & .867,~[.834, .899] & .881,~[.835, .926]\\
				\hline
			\end{tabular}
			\caption{Normalized FROC-AUC estimates for Test Set A and B.  Each entry shows a point estimate and a 95\% confidence interval.}
			\label{table:FAUC}
		\end{table}
		
		\subsection{Speed Evaluation}
		Because ESC systems are required to detect SPN-43 within a small time window, information about the speed at which each model performed detection is relevant. To test detection speed for each algorithm, a single data sample was loaded into memory along with any model parameters. The model was timed while performing detection on this single sample 200,000 times. The average detection time for each model on the single sample is listed in Table~\ref{table:timing}. Note that only the top-performing algorithms from each category are listed.

In order to ensure the timings were measured fairly, all measurements were performed on an Nvidia\textsuperscript{\textregistered} DGX\textsuperscript{TM} workstation. All timings were performed solely on the CPU during times when the workstation was not otherwise in use. While hardware specifics may change the values of the times listed in Table~\ref{table:timing}, we believe the relative ordering will remain consistent across most hardware platforms.

As expected, the sweep-integrated energy detector was faster than the machine learning methods.  CNN-3 was more than twice as fast as SVM$_{6164}$ and 4.5 times quicker than Inception-v1.  In the light of the results from the previous subsection, we can conclude that CNN-3 provides top-tier detection accuracy for significantly less computational cost than the other machine learning methods.   
	
		\begin{table}[htb]
			\centering
			\begin{tabular}{|r | c |}
				\hline
				Classifier & Time (ms) \\ \hline
				SI-ED & \textbf{0.40} \\
				CNN-3 & 1.89 \\
				SVM$_{6164}$ & 4.43 \\
				Inception-v1 & 8.45 \\
				\hline
			\end{tabular}
			\caption{The average detection time for each model on a single sample.}
			\label{table:timing}
		\end{table}

	\subsection{Detection Examples}
\begin{figure*}[!t]
			\centering	
			\includegraphics[width=.3\textwidth]{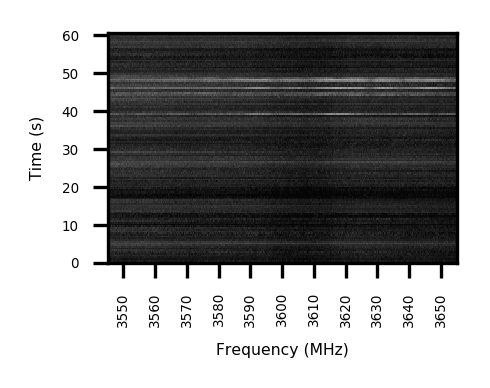}
			\includegraphics[width=.3\textwidth]{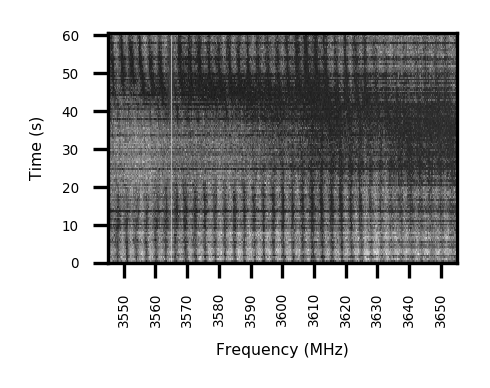}
			\includegraphics[width=.3\textwidth]{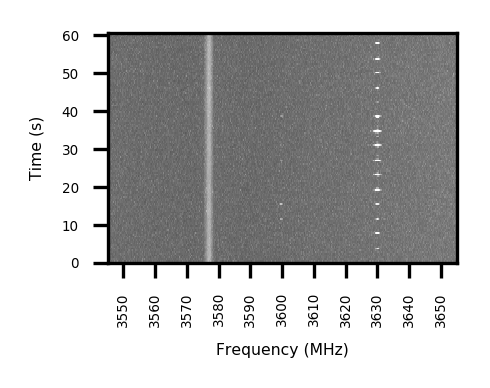}
			\caption{Example spectrogram captures, cropped to 3545-3655 MHz. (Top) Example 1: high-power Radar~3 OOBE; grayscale window [-95 -20]~dBm. (Middle) Example 2: low-power Radar~3 OOBE; grayscale window [-95 -70]~dBm. (Bottom) Example 3: SPN-43 emissions near 3630~MHz and a weak, barely visible SPN-43 at 3600 MHz; grayscale window [-100 -85]~dBm.}
			\label{fig:detection_examples}
		\end{figure*}

		To gain further insight into classifier performance, we examined spectrograms in which there was no consensus between machine learning and sweep-integrated energy detection.  For this analysis, as a representative machine learning algorithm, we chose CNN-3.  Both CNN-3 and sweep-integrated energy detection were applied with a decision threshold corresponding to a false-positive rate of 0.05 on Test Set~A.  Three notable examples of spectrograms in which sweep-integrated energy detection and CNN-3 differed, denoted Example~1, Example~2, and Example~3, respectively, are shown in Figure~\ref{fig:detection_examples}.

Example~1, shown in Figure~\ref{fig:detection_examples} (\ifbool{double_column}{Left}{Top}), contains high-power Radar~3 OOBE and no SPN-43 emissions.  In this case, sweep-integrated energy detection incorrectly detected SPN-43 in every 10~MHz channel, i.e., every channel was a false-positive.  On the other hand, CNN-3 had no false-positives.  

Example~2, shown in Figure~\ref{fig:detection_examples} (Middle), contains low-power Radar~3 OOBE and no SPN-43 emissions.  In this spectrogram, frequency-banding is evident, which may be due to multi-path fading.  For this example, energy detection correctly labeled all channels as negative for SPN-43.  However, CNN-3 incorrectly detected SPN-43 in several channels, resulting in false-positives.  This example illustrates a potential weakness of CNN-3 that could be explored in future work.

Last, Example~3, shown in Figure~\ref{fig:detection_examples} (\ifbool{double_column}{Right}{Bottom}), contains an evident SPN-43 emission at 3630~MHz and a faint SPN-43 emission at 3600~MHz.  The vertical line at 3577~MHz is local oscillator leakage, which appears bright due to the tight grayscale window.  In this example, sweep-integrated energy detection failed to detect both SPN-43 radars, which were correctly detected by CNN-3.

\section{Applications}
	\label{sec:Applications}
	Distributions estimated from field measurements for SPN-43 spectrum occupancy and SPN-43-absent power density are potentially informative to both federal regulators and commercial industry, as they may be relevant to ESC requirements \cite{WINNF-TS-0112,Sanders2017} and ESC development efforts.  Namely, occupancy distributions may be relevant to a requirement that the channel be vacated for a fixed time-interval after incumbent signals have been detected \cite{WINNF-TS-0112}.  Distributions of SPN-43-absent power may be relevant to ESC developers since some detection strategies may result in unacceptably high false-alarm rates for channels with higher levels of non-SPN-43 emissions.  In addition, field observations of ambient power levels are relevant to ESC certification testing \cite{Sanders2017}, since they could inform selection of background noise levels.  

As explained in Section~\ref{sec:spectrograms}, a total of 14,739 spectrograms were collected in San Diego and Virgina Beach, of which 4,491 were human-labeled for SPN-43 presence.  In this section, we describe how one of the best-performing classifiers from Section~\ref{sec:perf_eval}, CNN-3, was applied to classify the unlabeled spectrograms for SPN-43 presence and how the complete set of spectrograms was then used to estimate distributions for SPN-43 spectrum occupancy and for power density when SPN-43 was absent.  To classify unlabeled spectrograms for SPN-43 presence, we chose a decision threshold to generate the CNN-3 prediction output from each 10~MHz channel.  Below, we give details on the selected decision threshold, which was different for each application to accommodate dissimilar preferences between true-positive and false-positive rates.   

The findings given here are a partial selection from a larger set of descriptive statistics that is provided in an accompanying technical report \cite{TN2016}.  It should be emphasized that because these results are derived from spectrum observations at only two geographic locations for two months each, the reader should be careful not to draw overly-general conclusions.   

	\subsection{Channel Occupancy Statistics}
		For the goal of estimating SPN-43 channel occupancy, we chose the CNN-3 decision threshold to control the false-positive rate.  Specifically, the decision threshold was selected to correspond to a false-positive rate of $0.01$ on Test Set~A; this operating point corresponds to a true-positive rate of $0.97$.  Note that false-positives lead to a positive bias in occupancy estimates and a negative bias in vacancy estimates.  Thus, because we controlled the false-positive rate, our occupancy and vacancy estimates are conservative and liberal, respectively.  

As stated in Section~\ref{sec:spectrograms}, spectrograms were collected roughly every ten minutes.  This sampling interval was not exact due to hardware restrictions, like the rate at which data was saved to disk.  Despite this fact, to simplify our estimates of vacancy and occupancy time-intervals, we assumed that the captures were exactly ten minutes apart.  To calculate the length of time a 10~MHz channel was either occupied by SPN-43 or vacant, we ordered the spectrograms by their capture time-stamp and then counted the number of consecutive vacant and occupied observations. The counts were then multiplied by 10 minutes to estimate durations.  Note that this approach could not resolve changes in SPN-43 occupancy that occurred less than 10 minutes apart.  

\begin{figure}[tb]
			\centering
			\includegraphics[width=4.3cm]{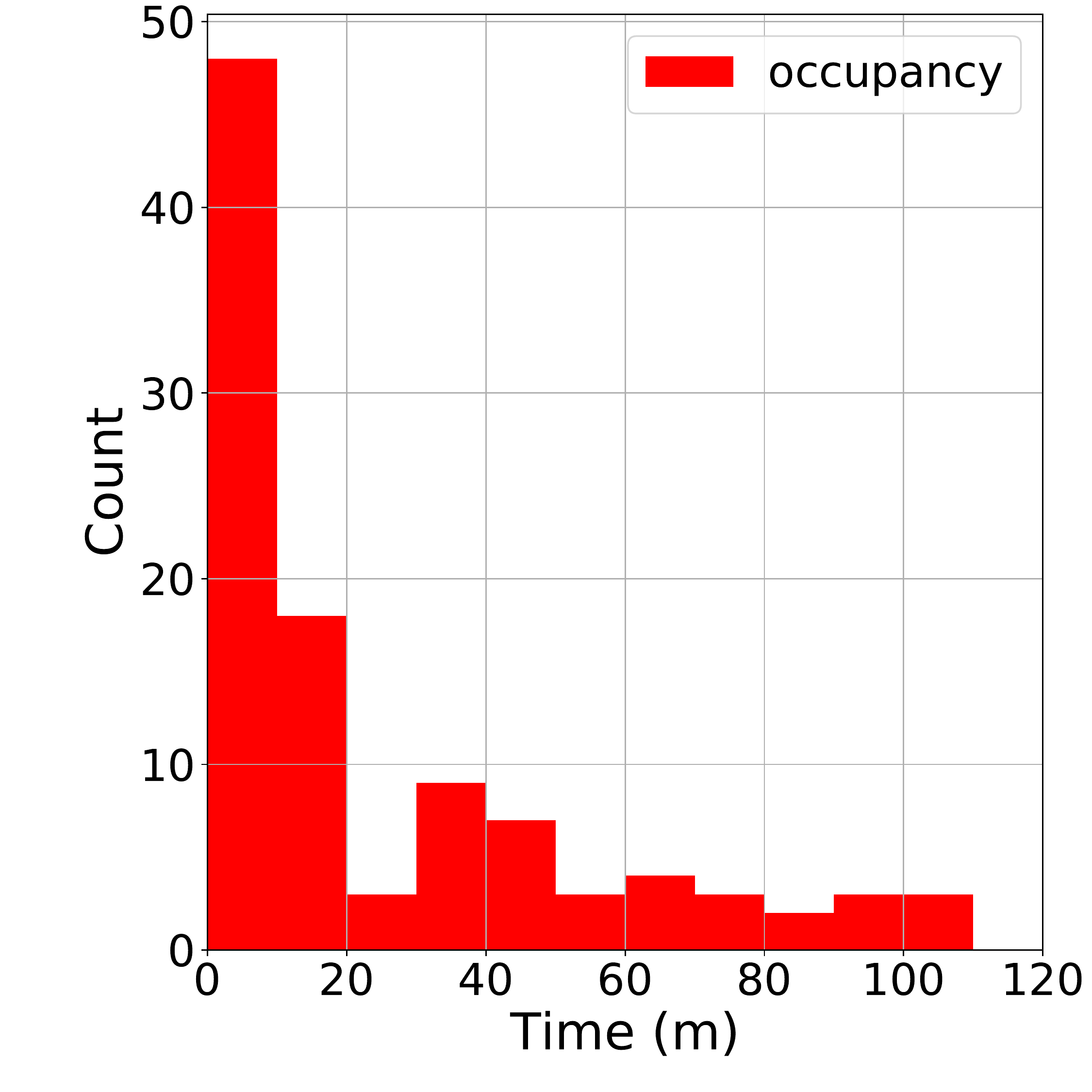}
			\includegraphics[width=4.3cm]{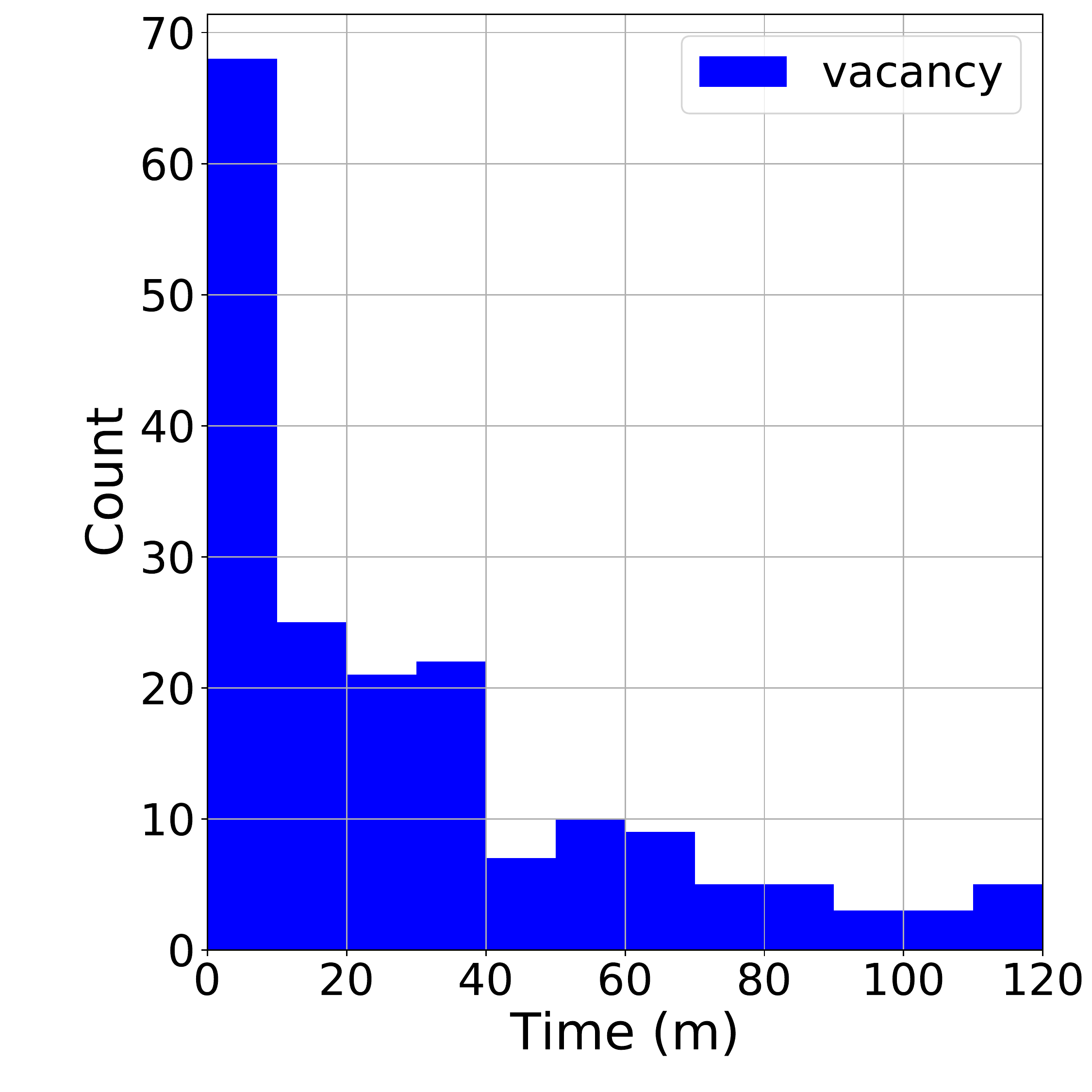} 
			\caption{Histograms of the empirical length of time the 10~MHz channel centered at 3550~MHz was occupied (left) or unoccupied (right) by SPN-43 in San Diego for time intervals less than 120 minutes.}
			\label{fig:occupancy_hist}
		\end{figure}

Figure~\ref{fig:occupancy_hist} shows histograms of occupied and vacant time-intervals measured in minutes for the 10~MHz channel centered at 3550~MHz in San Diego.  Specifically, the occupancy histogram lists the number of time-intervals for which SPN-43 was continuously-observed for the specified duration.  For example, the 3550~MHz channel was continuously occupied for 30-40 minutes nine times during the two-month measurement period in San Diego.  Similarly, the vacancy histogram lists the number of time-intervals for which SPN-43 was not present for the specified duration, e.g., there were ten SPN-43 vacancies with durations of 50-60 minutes.  Only time-intervals below 120 minutes are shown.  Of the observed time-intervals, 24 occupancies and 138 vacancies exceeded 120 minutes.  

To gain a better understanding of how often a channel was occupied, we estimated the occupancy ratio, i.e., the amount of time the channel was occupied by SPN-43 divided by the total observation time.  Table~\ref{table:occ_ratio} lists the estimated occupancy ratio for channels where SPN-43 was observed in San Diego and Virgina Beach, respectively.  

		\begin{table}[tb]
			\centering
			\begin{tabular}{| r | c | c |}
				\hline
				Measurement Site & Channel Center (MHz) & Occupancy Ratio \\ \hline
				San Diego & 3520 & $0.036 \pm 0.002$ \\
				San Diego & 3550 & $0.287 \pm 0.005$ \\
				San Diego & 3600 & $0.018 \pm 0.001$ \\
				Virginia Beach & 3570 & $0.116 \pm 0.004$ \\
				Virginia Beach & 3600 & $0.118 \pm 0.004$ \\
				Virginia Beach & 3630 & $0.041 \pm 0.003$ \\
				\hline
			\end{tabular}
			\caption{Estimated occupancy ratios for 10~MHz channels in which SPN-43 was observed.  The error bands indicate 95\% confidence intervals.  Note that SPN-43 was observed at 3520~MHz in San Diego \cite{TN1954}, which is outside of the CBRS band.}
			\label{table:occ_ratio}
		\end{table}

	\subsection{Power Density Distributions}
		For the aim of estimating the distribution of power density when SPN-43 was absent, we chose the CNN-3 decision threshold to control the number of missed SPN-43 detections (false-negatives).  Specifically, the decision threshold was selected to correspond to a true-positive rate of $0.98$ on Test Set~A; this operating point corresponds to a false-positive rate of $0.02$.  Because missed detections lead to the inclusion of SPN-43 emissions in estimates of the SPN-43-absent power density, and because the power density of SPN-43 emissions is typically quite high, missed detections are expected to add a positive bias.  Therefore, to avoid such a bias, we controlled the rate of missed detections with the potential expense of additional false-positives, which shrank our sample size for SPN-43-absent observations.    

\begin{figure}[tb]
			\centering
			\includegraphics[width=4.3cm]{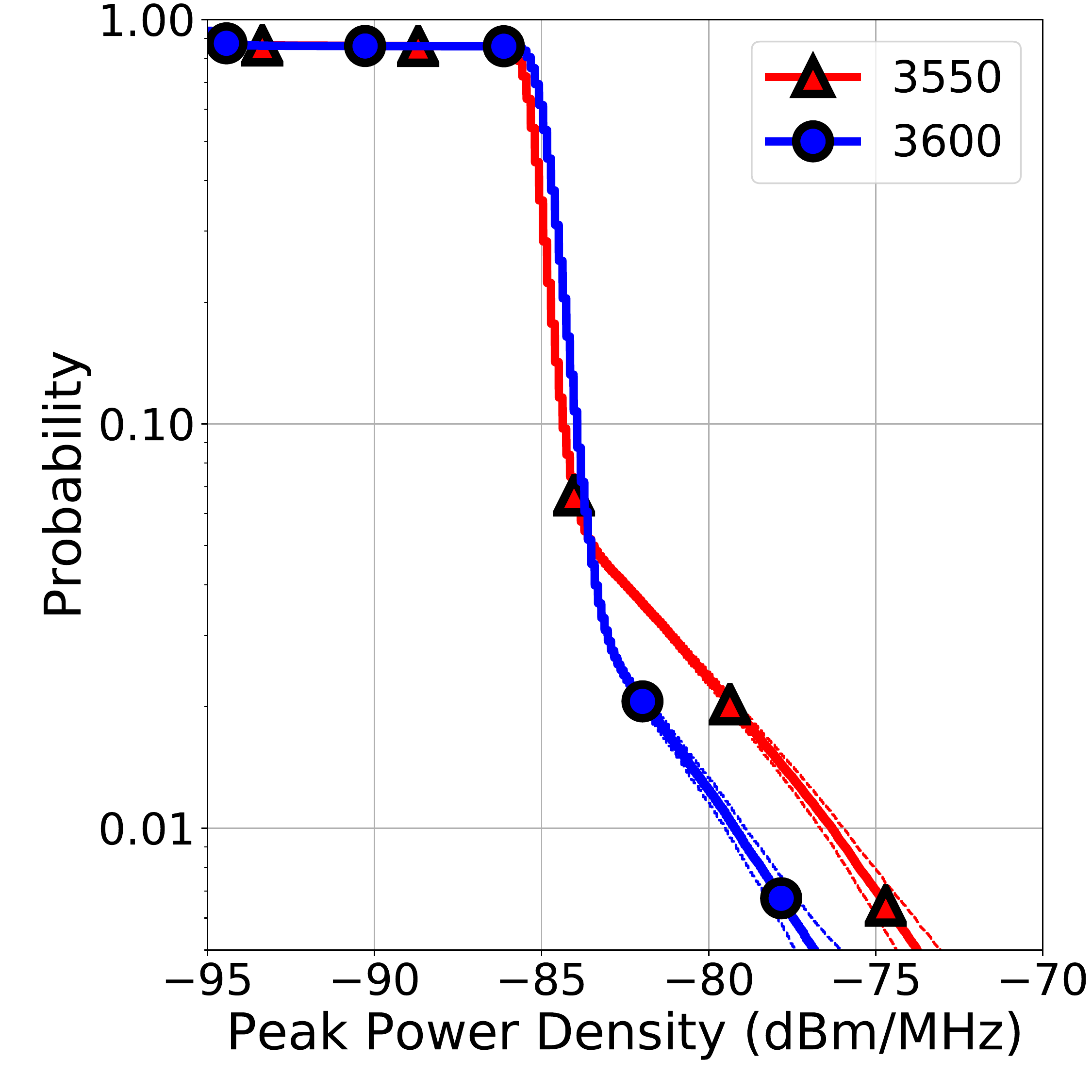}
			\includegraphics[width=4.3cm]{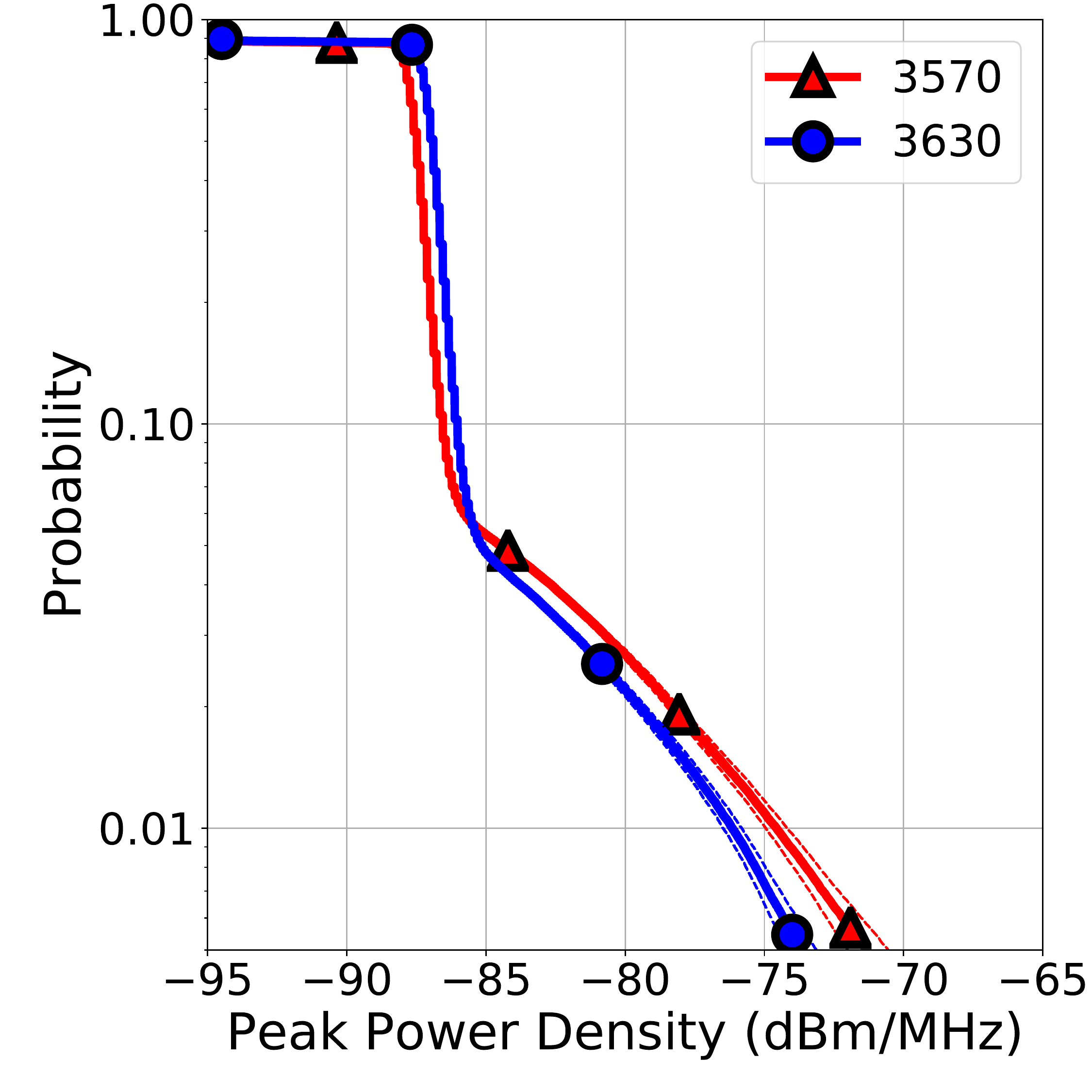}
			\caption{The power CCDF for emissions captured using the CBS antenna when SPN-43 was absent in San Diego (Left) and Virginia Beach (Right). Simultaneous 95\% confidence bands are indicated by dashed lines.}
			\label{fig:power_density_ccdf}
		\end{figure}

After classification of the unlabeled spectrograms, the channels found to contain SPN-43 were discarded, and the empirical cumulative distribution function (CDF) for the power density was estimated from the set of spectrogram values (converted to dBm/MHz) in the 220~kHz-wide frequency-bin at the center of each 10~MHz channel (the expected location for SPN-43).  Figure~\ref{fig:power_density_ccdf} shows examples of the empirical complementary CDF (CCDF), equal to one minus the CDF, for the SPN-43-absent power density for frequencies where SPN-43 was observed in San Diego and Virgina Beach.  These plots can be used to quickly read off percentiles associated with the upper tails of each distribution.  Namely, the 90th and 99th percentiles correspond to the power density values where the CCDF is equal to 0.1 and 0.01, respectively.  Nonparametric simultaneous 95\% confidence bands were estimated for the empirical CCDFs using a method based on the Dvoretzky-Kiefer-Wolfowitz inequality \cite[Thm. 7.5]{Wasserman2004}.  The steep decrease in the CCDFs corresponds to the noise floor of the receiver, which varied with measurement location and receiver reference level \cite{TN1954,TN1967,TN2016}.  The tail heaviness indicates the prevalence of non-SPN-43 emissions, such as Radar~3 OOBE. 

\section{Summary and Discussion}
	Accurate detection of radar in the 3.5~GHz band with ESC sensors is needed to both protect federal incumbent systems and to enable economical commercial utilization of the band.  In this paper, for the task of single-channel SPN-43 radar detection from spectrograms collected with a single receiver, we investigated the effectiveness of thirteen detection algorithms, including eight deep learning methods, three classical machine learning approaches, and two energy detection strategies.  Furthermore, for the task of wideband SPN-43 detection across multiple channels observed concurrently with one receiver, we compare the top-performing methods from the single-channel evaluation.  The detection algorithms were trained and tested with a set of nearly 4,500 unverified, human-labeled spectrograms collected at two coastal locations.  

For ESC networks to be effective, they will require a large numbers of detectors. Detection algorithms that are fast and computationally inexpensive are ideal for such networks because they reduce the costs for the large number of individual sensors.  For this reason, we chose to explore the use of spectrograms for detection, as both generating and processing spectrograms requires relatively inexpensive hardware.  Algorithms for detecting emissions from other data representations such as I/Q recordings may result in higher detection accuracy but also require more expensive hardware to generate and process.  Nonetheless, the use of I/Q recordings is an area for potential future work.

Our evaluations provide estimates for the level of performance that can potentially be expected from an ESC detector across a range of real-world conditions. In particular, our test set included cases with realistic channel conditions and Radar~3 OOBE.  Despite these advantages, there are several drawbacks to using real data.  First, using real-world data with unknown signal and noise components prevents accurate assessment of detection performance as a function of SNR.  This type of analysis can provide information about emission strength that must be present for detectors to succeed.  In future work, we plan to evaluate detection strategies using synthetic data generated from our field measurements in which the SNR can be controlled.  

Second, capturing a desired ratio of observations for different subgroups is difficult in the field.  Consequently, our dataset had a limited number of cases available to construct training, validation, and test sets with sufficient representation of the subgroups listed in Table~\ref{test_set_composition}.  To ensure that our training and testing sets contained enough cases for each relevant subgroup, we chose to not use a separate validation set \cite[Sec.~1.4.8]{Murphy2012}.  Instead, in an attempt to avoid over-fitting, we stopped training the deep learning models after a fixed number of epochs.

Our evaluations on real-world data demonstrated that machine learning methods offer superior detection performance compared to methods based on energy detection.  The excellent detection performance of CNN-3 allowed us to estimate spectrum occupancy statistics and power distributions for non-SPN-43 emissions with a much higher accuracy than would have been otherwise possible.  In particular, using energy detection to classify the unlabeled data would have resulted in many more false-positives and missed detections at a given decision threshold, which would lead to biased estimates.  Namely, false-positives lead to overestimation of spectrum occupancy, whereas missed detections result in an overestimation of spectrum vacancy and add a positive bias to non-SPN-43 power distributions.  As explained in Section~\ref{sec:Applications}, occupancy statistics and power distributions may have value to both ESC developers and spectrum regulators.  A complete set of occupancy statistics and power distributions for each channel in the 3.5~GHz band is provided in a technical report \cite{TN2016}.

\appendices
	\section{ROC and FROC Curves}
\label{ROCappendix}

We present a brief introduction to two types of graphical plots that can be used to evaluation detection performance: the receiver operating characteric (ROC) curve and a related generalization, called the free-response ROC (FROC) curve.  Further background on ROC curves can be found in \cite{VanTrees1968,Metz1978,Pepe2003} and details on FROC curves are given in \cite{Bunch1978,Metz2006}.  Although ROC curves are commonly utilized in signal processing and machine learning, FROC curves are lesser-known, since they have been primarily applied in radiology to evaluate lesion detection performance.  In the context of multichannel spectrum sensing for cognitive radio, a notable application of FROC curves to classifier performance evaluation is the work of Collins and Sirkeci-Mergen \cite{Collins2013}.

\subsection{Binary Signal Detection: The ROC Curve}
For a binary signal detection task, the aim is to use a data observation to decide whether or not a signal is present.  Each decision results in one of four possible outcomes: true-positive (TP), false-positive (FP), true-negative (TN), or false-negative (FN).  These outcomes give rise to four conditional probabilities (or rates).  In the engineering literature,  the TP rate, FN rate, and FP rate are commonly called the ``detection'', ``miss'' and ``false-alarm'' probabilities, respectively.  For a given decision threshold, binary classification performance is fully described by the FP and TP rates.  Namely, the FN rate is equal to one minus the TP rate, and the TN rate is one minus the FP rate.  

\begin{figure}[tb]
	\centering
	\includegraphics[width=4.3cm]{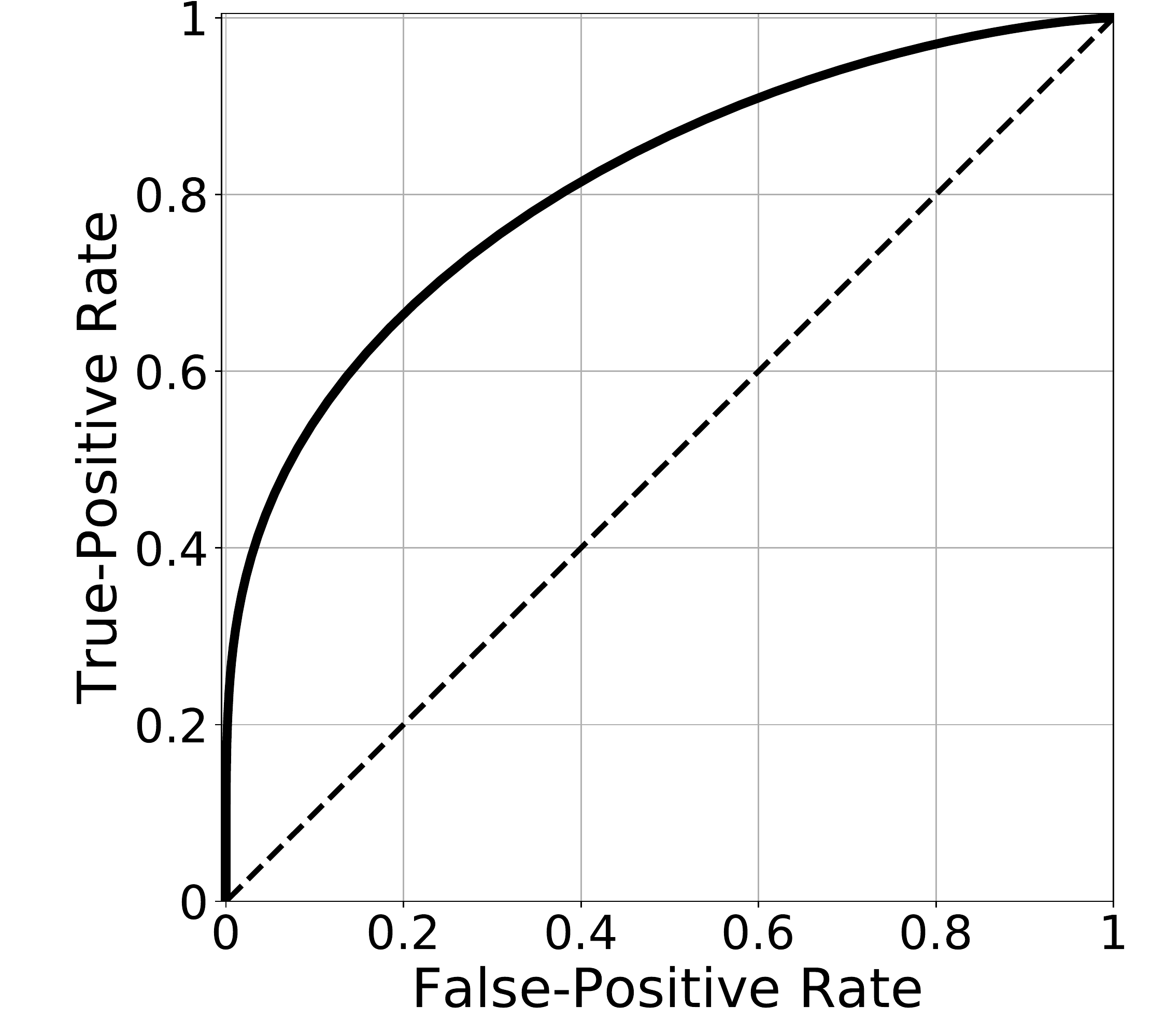} 
	\includegraphics[width=4.3cm]{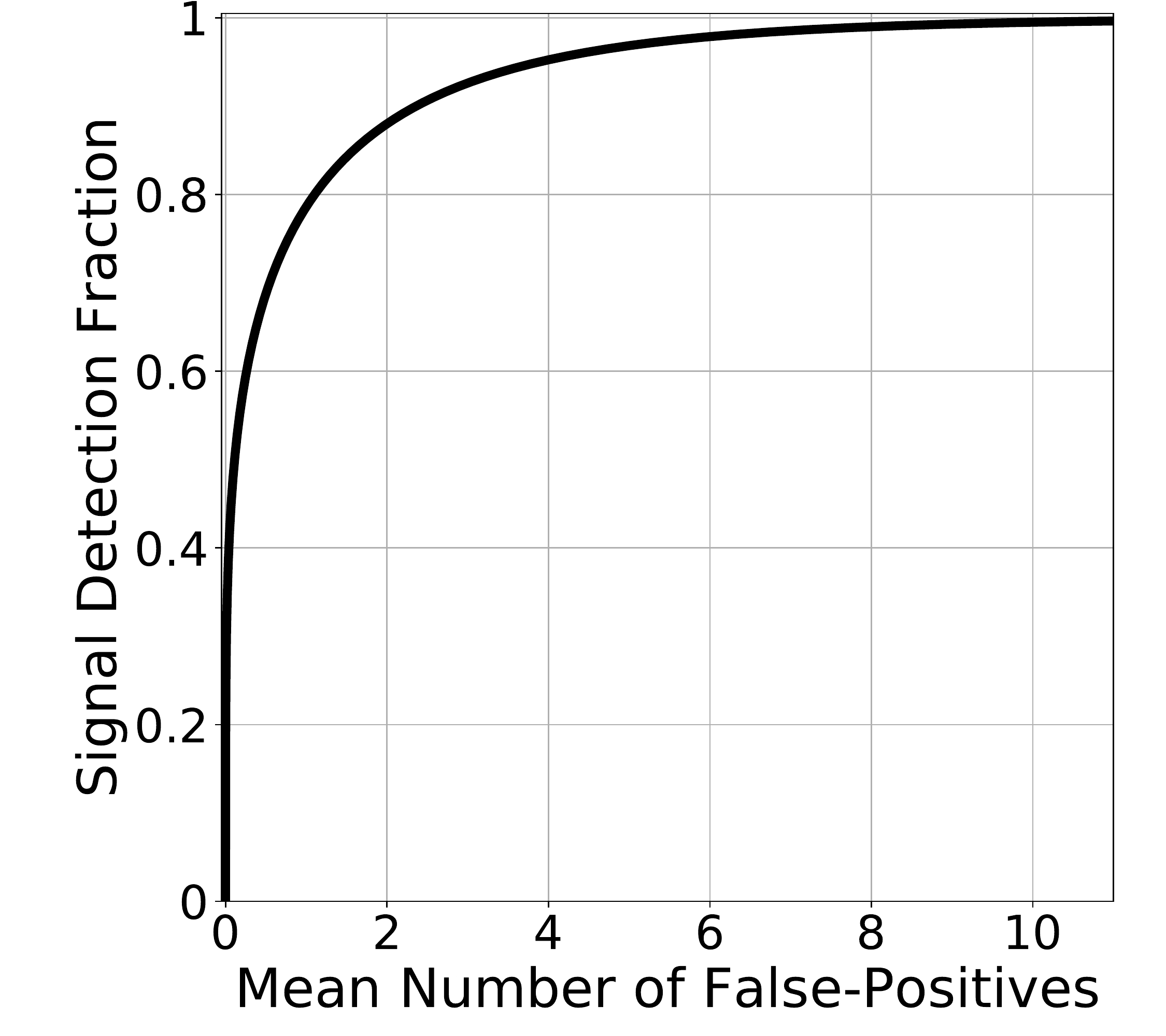}
	\caption{Examples of an ROC curve (Left) and an FROC curve (Right).  In the ROC plot, the ``chance line'' is depicted with the diagonal dashed line.}
	\label{fig:ROCexamples}
\end{figure}

A useful way to summarize detection performance is the ROC curve, defined as the plot of TP rate versus FP rate, over all decision thresholds \cite{VanTrees1968,Pepe2003}.  An example of an ROC curve is shown in Figure~\ref{fig:ROCexamples} (left).   When comparing ROC curves, better classifier performance is indicated by a higher curve that is closer to the upper left corner.  Namely, for a perfect classifier, there exists a threshold where the TP rate is equal to one with a FP rate of zero.  By contrast, for a useless classifier, the ROC curve is equal to or below the diagonal dashed ``chance line'' shown in Figure~\ref{fig:ROCexamples} (left) for which the TP rate is equal to the FP rate at all decision thresholds \cite{Pepe2003}.  

ROC curves possess three properties that make them particularly useful.  First, they fully characterize binary classifier performance over all decision thresholds, which enables evaluation and comparison of classifiers that may be deployed at various operating points (thresholds) \cite{Metz1978}.  Second, ROC curves are invariant under strictly-increasing transformations of the decision variable \cite{Pepe2003}.  Thus, classifiers with decision variables on different ordinal scales can be compared via ROC curves.  Third, ROC curves are independent of signal prevalence \cite{Metz1978}.  This implies that ROC curves can be used to assess classifiers that may be deployed in environments with different signal prevalences.         

A commonly-used summary measure for binary classification performance is the area under the ROC curve (ROC-AUC).  ROC-AUC takes values between zero and one, with higher values indicating better performance.  ROC-AUC can be interpreted as the average TP rate, averaged uniformly over all FP rates.  Alternatively, ROC-AUC can be interpreted as a probability.  Namely, given randomly-selected signal-absent and signal-present cases, ROC-AUC is the probability that the signal-present case is rated higher \cite[Sec. 4.3]{Pepe2003}.  

In this paper, we use the so-called ``empirical'' nonparametric estimators for the ROC curve and its area.  For details on these estimators, see \cite[Sec. 5.2]{Pepe2003}.  In addition to having simple implementations, the empirical estimators are nonparametric and unbiased.   

\subsection{Multiple Signal Detection and Localization: The FROC Curve}
The FROC curve \cite{Bunch1978,Metz2006} is generalization of the ROC curve designed to summarize classifier performance for a combined detection and localization task in which multiple detection decisions are made for each observation.  An example of such a task arises in multichannel spectrum sensing, where the aim is to detect one or more signals and localize them in frequency.  After specifying a criterion for correct signal localization, it is possible to determine if a detection result is a correctly-localized TP or a FP.  Suppose that each correctly-localized TP detection occurs with the same probability, called the signal-detection fraction.  The FROC curve is defined as the plot of signal detection fraction versus the mean number of false-positives per observation, plotted over all decision thresholds; an example FROC curve is shown in Figure~\ref{fig:ROCexamples}.  Like the ROC curve, an FROC curve closer to the upper left corner of the graph indicates better classifier performance.  To estimate the FROC curve, we use the usual empirical estimator \cite{Bunch1978}.

When the number of detection decisions is bounded, the abscissa (x-axis) of the FROC curve is bounded, and the area under the FROC curve (FROC-AUC) is a well-defined summary measure.  For example, multichannel spectrum sensing typically aims to assess spectrum occupancy for a fixed number of frequency channels.  Because the maximum abscissa value for an empirical FROC curve depends on the maximum number of possible FP decisions in the test set, the maximum empirical FROC-AUC can be different for dissimilar test sets.  Therefore, in this paper, to enable straightforward comparisons between test sets, we normalize FROC-AUC values to fall between zero and one.   The normalization factor depends on the maximum number of possible FP decisions in the test set.

\subsection{When to Use Which Curve?}

Because ROC and FROC curves are designed for different, but related tasks, they provide complementary insights into classifier performance.  In particular, ROC curves focus solely on signal detection for a single decision, regardless of signal localization.  Thus, for the problem of multichannel spectrum sensing, ROC curves are best suited to low-level assessment of classifier performance for a single channel.  Such evaluations may be particularly useful for classifier development.  On the other hand, FROC curves assess both detection and signal localization when multiple decisions must be made.  For this reason, they are better suited to classifier performance evaluation for the full multichannel spectrum sensing task.  If FROC curves are not a good match to the task and associated preferences, one can consider variations of FROC curves that weight TP and FP decisions differently; for further details on FROC variants and their generalizations, see \cite{Wunderlich2016}.

\bibliographystyle{ieeetr}
\bibliography{citations}

\end{document}